\documentclass[a4paper,fleqn,usenatbib,useAMS]{mnras}
\usepackage{graphicx}	
\usepackage{amsmath}	
\usepackage{multicol}        
\usepackage{multirow}

\title[Dust evolution]{The role of dust destruction and dust growth in the evolution of the interstellar medium}
\author[O. Osman, K. Bekki and L. Cortese]
{Omima Osman ${}^{1,}$ ${}^2$\thanks{E-mail:
omima.osman@icrar.org},
Kenji Bekki${}^1$ and Luca Cortese${}^1$\\ 
${}^1$ICRAR M468
The University of Western Australia
35 Stirling Hwy, Crawley
Western Australia 6009, Australia\\
${}^2$
University of Khartoum - Department of Physics.
Al-Gamaa Ave, Khartoum 11115, Sudan.
P.O.Box 321}

\begin{document}

\date{Accepted, Received 2005 February 20; in original form }

\pagerange{\pageref{firstpage}--\pageref{lastpage}} \pubyear{2005}

\maketitle

\label{firstpage}

\begin{abstract}
We use Milky Way--like chemodynamical simulations with a new treatment for dust destruction and growth to investigate how these two processes affect the properties of the interstellar medium in galaxies. We focus on the role of two specific parameters: $f_{des}$ (a new parameter that determines the fraction of dust destroyed in a single gas particle surrounding supernova) and $C_s$ (the probability that a metal atom or ion sticks to the dust grain after colliding, i.e., the sticking coefficient) in regulating the amount and distribution of dust, cold gas and metals in galaxies. We find that simulated galaxies with low $f_{des}$ and/or high $C_s$ values produce not only more dust, but they also have a shallower correlation between dust surface density and total gas surface density, and a steeper correlation between dust--to--gas ratio and metallicity. Only for values of $f_{des}$ between 0.01 and 0.02, and of $C_s$ between 0.5 and 1 our simulations produce an average slope of the dust--to--gas ratio versus metallicity relation consistent with observations. $f_{des}$ values correspond to a range of a total fraction of dust destroyed by a single supernova between 0.42 and 0.44. Lastly, we compare predictions of several simulations (with different star formation recipes, gas fractions, central metallicities, and metallicity gradients) to the spatially resolved M101 galaxy, and conclude that metallicity is the primary driver of the spatial distribution of dust, while dust--to--gas ratio controls the cold gas distribution, as it regulates the atomic--to--molecular hydrogen conversion rate.

\end{abstract}

\begin{keywords}
ISM:dust --
galaxies:ISM --
galaxies:evolution --
stars:formation  
\end{keywords}

\section{Introduction}
Interstellar dust represents a small fraction of the total mass budget of galaxies, about 0.01 of the interstellar medium (ISM) and 0.001 of the stellar mass (Draine 2007; Smith et al. 2012). Despite its small amount, it has a significant influence on the spectral energy distribution of galaxies (SED) through absorption and scattering of the ultraviolet (UV) radiation and re--emission of the far--infrared and submillimeter photons (see Draine 2003 and Galliano et al. 2018 for reviews). Accordingly, when calculating quantities such as star formation rate (SFR) from observed fluxes, a dust correction has to be applied. Dust also forms a catalyst on which molecules form, in particular, molecular hydrogen forms (H$_{2}$; Wakelam et al. 2017) and hence supplies star formation with its necessary fuel (e.g., Fukui \& Kawamura 2010). Cazaux \& Tielens (2004) found that for temperatures $\leq$ 20 K, the efficiency of H$_{2}$ formation on dust grains is near unity and is highly efficient at high temperatures too. More generally, dust affects the chemistry of the ISM by affecting the chemical pathways of a number of the ISM molecules that form on its surface (Dulieu et al. 2013).

Our understanding of dust formation and evolution has evolved through decades of study. Asymptotic giant branch (AGB) stars and supernovae (SNe) are widely accepted as the primary sources of what is called `Stardust'. AGB stars supply the ISM with either oxygen--rich or carbon--rich dust depending on the ratio of carbon to oxygen in their atmospheres (e.g. Sargent et al. 2010; Srinivasan et al. 2010), contrarily to SNe which supply the ISM with both types of dust (Dwek 1998; hereafter D98). Theoretical studies indicate that SNe produce 0.1--1 M$_{\sun}$ of dust (Bianchi \& Schneider 2007; De Looze et al. 2017), while AGB stars produce $10^{-3}$--$10^{-2}$ M$_{\sun}$ (Ferrarotti \& Gail 2006; Ventura et al. 2012a; 2012b; Dell' Agli  2017a; 2017b) with a significant dependence on the stellar metallicity.
\begin{table*}
\caption{Summary of recent literature on dust physics using hydrodynamical simulations. DM, AGN, and PEH stand for dark matter, active galactic nuclei, and photoelectric heating, respectively.}
\begin{tabular}{|lll|}
\hline
{Paper}
& {New physics input in the simulations}
& {Summary}\\
\hline
{\multirow{1}{*}{Bekki 2013}} & Dust formation, growth, destruction, & Star formation histories of disc galaxies are regulated by dust evolution\\
 & and H$_2$ formation on dust. & since H$_2$--dependent star formation is controlled by dust properties.\\[0.2cm] 

{\multirow{1}{*}{Bekki 2015}} &  Treating dust as separate component & Radiation pressure of stars plays important role in dust distributions and\\
& of the galaxy (DM, stars, gas, and dust). & accordingly in gas distributions, chemical abundance and H$_2$ fraction as well.\\[0.2cm] 

{\multirow{1}{*}{McKinnon et al. 2016}} & Dust formation, growth, and destruction  & Both Stellar and AGN feedback are important in reproducing the observed \\ 
 & in cosmological zoom--in simulations &  dust--metallicity relation.  At redshift ($z$) = 0 dust originated from SNe\\ 
 & (implemented in the moving--mesh  & type II dominates dust originated from AGB stars.\\
 & code AREPO). & \\[0.2cm] 

\multirow{1}{*}{Zhukovska et al. 2016} & Dust formation, growth, and destruction & The dust sticking coefficient decreases with the gas temperature.\\
 & in multiphase, inhomogeneous ISM. & \\[0.2cm] 

\multirow{1}{*}{Forbes et al. 2016} & Time and space varing PEH in dwarf & PEH feedback dominates SNe feedback (but see Hu et al. 2017).\\
& galaxies. & \\[0.2cm] 

\multirow{1}{*}{Aoyama et al. 2017} & Grain shattering, coagulation, and grain & Grain growth by accretion is triggered by shattering and becomes active \\
 & size distribution (presented by two   & when the galaxy is of age of 0.2 Gyr. Coagulation becomes significant when  \\
 & populations, small and large grains).   &the abundance of small grains is high enough at $\sim$ 1 Gyr.\\[0.2cm] 

\multirow{1}{*}{Hou et al. 2017} & Grain shattering, coagulation, grain size & Extinction curves evolve consistently with dust evolution, being flat with no\\
 & distribution (presented by two & 12175  $\AA$ bump in the early evolution stages (t $\leq$ 0.3 Gyr) due to the\\
 & populations, small and large grains), and  & domination of large grains produced by stars. Then they become\\
 & chemical composition of dust grains. & steeper with prominent 2175 $\AA$ bump when grain growth and shattering \\
 & & become efficient.\\[0.2cm] 

\multirow{1}{*}{McKinnon et al. 2018} &  Dust grains dynamics and size evolution. & Presented  a  novel numerical framework for the dynamics and size evolution\\
 & & of dust grains and demonstrated the importance of shattering in \\
 & & steepening extinction curves.\\[0.2cm] 

\multirow{1}{*}{Gjergo et al. 2018} & Grain shattering, coagulation, grain size & At $z$ $\geq$ 3 proto--cluster regions are already rich in dusty gas and the\\
 & distribution (presented by two, &   dust properties at this stage is significantly different from those observed\\
 & populations, small and large grains),  & in the Milky Way.\\
 & and chemical composition of dust grains & \\
 & in galaxy clusters. & \\[0.2cm] 

\multirow{1}{*}{Hu et al. 2019} & Dust dynamics in turbulent multiphase & Dust destruction timescale ($\tau$) is 0.35 Gyr for silicate dust and 0.44 Gyr\\
 &  ISM. &  for carbon dust in solar neighbourhood conditions and they\\
 &  & are dependent on the SN environment.\\

\hline
\end{tabular}
\end{table*}

Dust grains experience many destruction processes caused by SNe such as thermal and non--thermal sputtering. Dust produced by SNe passes through the shocked region before arriving in the ISM, in this passage, most of the small dust grains are destroyed. As a result, the grain size distribution is biased towards large grains, 0.1 $\micron$. Grains produced by AGB stars are also biased towards 0.1 $\micron$ (Gauger et al. 1999; Kozasa et al. 2009; Yasuda \& Kozasa 2012). In the low--density ISM, dust destruction by SNe remnants continues in a timescale of a few hundreds of Myrs (Barlow 1978; Mckee 1989; Jones et al. 1994; Nozawa et al. 2006; Yamasawa et al. 2011; Andersen et al. 2011). This timescale depends on the total mass of the ISM, the mass of the ISM swept up by SNe (M$_{swept}$), the destruction efficiency of SNe, and SNe rate. Nozawa et al. (2006) considered M$_{swept}$ dependency on the gas density and found that the lower the gas density is, the higher the  M$_{swept}$ is,  and hence the process is more efficient. They also found that SNe destruction efficiency increases as a function of the explosion energy (E$_{sn}$) and/or the ISM gas density, and has a dependence on the initial size distribution of dust and the sputtering yield.

Mckee (1989) reported that only about 20\% of the dust produced could survive destruction processes. This fraction does not account for the dust amounts observed in low and high redshift galaxies (Jones et al. 1994; Bertoldi et al. 2003; Mattsson 2011; Kuo \& Hirashita 2012; McKinnon et al. 2016; Ginolfi et al. 2018; Aoyama et al. 2018; McKinnon et al. 2018). Dust growth in the ISM by accretion of the gas--phase metals onto pre--existing dust cores is assumed to solve the puzzle of the missing dust (Dwek \& Scale 1980). Dust growth is most efficient in dense neutral and molecular clouds with density $n_H$ $\approx 10^{2}-10^{3} cm^{-3}$ (Hirashita 2000). Inoue (2011) and Asano et al. (2013) argued that there is a critical metallicity above which the contribution of dust growth overcomes the contribution of stellar sources to the dust mass. Since accretion is a surface process, it is highly dependent on the grain size distribution. Small grains are favoured in the process because of their higher surface to volume ratios (Hirashita \& Kuo 2011). Accordingly, this process changes the total dust mass and the grain size distribution. Zhukovska et al. (2016; 2018) concluded that dust growth is essential to explain the depletion levels of silicon and iron in the Milky Way.

The balance between dust destruction processes and dust growth determines the dust mass in an isolated galaxy to a large extent. In this framework,  parameters such as dust destruction efficiency ($F_{des}$) and dust sticking coefficient ($C_s$: the probability that a metal atom or ion sticks to the dust grain after colliding) affect dust abundance significantly. $C_s$ is mainly taken to be unity in literature with a few exceptions that adopted lower values (e.g. Hirashita 2000; Hirashita \& Kuo 2011; Aoyama et al. 2017), while $F_{des}$ is taken to be 0.1 following Mckee (1989) and Nozawa et al. (2006) (e.g. Asano et al. 2013). McKinnon et al. (2016) adopted $F_{des}$ of 0.3.  Table 1 contains a summary of recent literature that used hydrodynamical simulations to study dust. Despite the commonly used values of $F_{des}$ and $C_s$ in literature, $F_{des}$ and $C_s$  vary according to the dust properties and composition, and the properties of the ambient medium (e.g., Jones et al. 1994; Serra Diaz--Cano \& Jones 2008; Bocchio et al. 2014; and references therein).

Thus, the purpose of this paper is to study the influence of the variations of dust parameters on the dust--to--gas ratio versus metallicity relation, and dust and gas spatial correlations. We also attempt to constrain the possible range of those parameters. To this end, we use a chemodynamical simulation that takes into account the dependence of the destruction and growth processes on the ISM gas properties such as density, temperature, and metallicity. The simulation used in this study can track the processes mentioned above and H$_2$ formation on dust grains, i.e. can predict simultaneously dust and H$_2$ properties which are not possible in other simulations (e.g. Krumholz, McKee \& Tumlinson 2009; Fu et al. 2010; Kuhlen et al. 2012). Furthermore, we introduce a new parameter $f_{des}$. $f_{des}$ is a parameter that determines the fraction of dust destroyed in a single SPH particle surrounding SN, i.e. in our simulation, SNe effect on dust extends to the neighbouring gas particles (see section 2.2 for more details). In the rest of this paper, we focus on studying and constraining $f_{des}$ and $C_s$ referred to as dust parameters.

Despite the improvements mentioned above, our simulation has limitations regarding the dust physics implemented. For instance, we do not include shattering, grains size distribution, and grains composition. Additionally, it is difficult to resolve ISM densities where a significant fraction of the dust growth occurs (n$_H \approx 10^{2}-10^{3} cm^{-3}$; Hirashita 2000). To address this problem Hirashita \& Aoyama (2019) and  Aoyama et al. (2017),  for instance, switched on dust growth in gas particles with densities higher than 10 $cm^{-3}$ and temperatures below 10$^3$ K, this gas is assumed to host the unresolved dense gas where growth occurs. In the present simulation, the resolution of the gas particles is not enough as well (m$_g = 3\times10^4 {\rm M}_{\odot}$; see Table 2 for full parameter description).  Although a small fraction (1\%) of the ISM in our simulation lies in the range identified by Hirashita (2000) for the characteristic density for dust growth, dust growth in our simulation is switched on in all gas particles. In this case, dust growth occurs throughout the ISM; however, dust grains can hardly grow in low--density environments and vice versa. Effects of the simulation resolution are addressed in section 5.

To tackle the tasks of this paper, we used the dust--to--gas ratio versus metallicity ($log(D/G)$--$[12+log(\frac{O}{H})]$), H$_2$ and total gas surface densities versus dust surface density ($\Sigma_{H_2}$--$\Sigma_D$, $\Sigma_{G}$--$\Sigma_D$) relations.  While $log(D/G)$--$[12+log(\frac{O}{H})]$ provides insights into the time evolution of the dust content and its processing in the ISM (Leroy et al. 2011; Galliano et al. 2011; Sandstrom et al. 2013;  Remy--Ruyer et al. 2014; Aoyama et al. 2017; Roman-Duval et al. 2017; Relano et al. 2018), $\Sigma_{H_2}$--$\Sigma_D$ and $\Sigma_{G}$--$\Sigma_D$ give insights into dust spatial correlations. In particular, $\Sigma_{H_2}$--$\Sigma_D$ provides information about the spatial relationship between the two and the interplay between H$_2$ formation enhanced by dust and dust growth enhanced in molecular clouds. $\Sigma_{G}$ is observed to be well correlated with $\Sigma_D$ (e.g. Leroy et al. 2011). The combination of those relations allows us to use the temporal and spatial properties of the dust to make our predictions. 

We also attempt to compare predictions of the simulation to the data of the spatially resolved M101 galaxy. M101 is a galaxy with a wealth of observational data and a large metallicity gradient that ranges from 7.8 to 8.7 in oxygen abundance units (Croxall et al. 2016). This wide range of metallicity is representative of a variety of environments which makes M101 galaxy useful for testing our simulation. The rest of this paper is organised as follows: description of the chemodynamical simulation is given in section 2, the main results are displayed in section 3. Sections 4 and 5 contain the comparison with M101 galaxy and the discussion and conclusions of the study.  

\begin{figure*}
\begin{multicols}{2}
    \includegraphics[width=1.\linewidth]{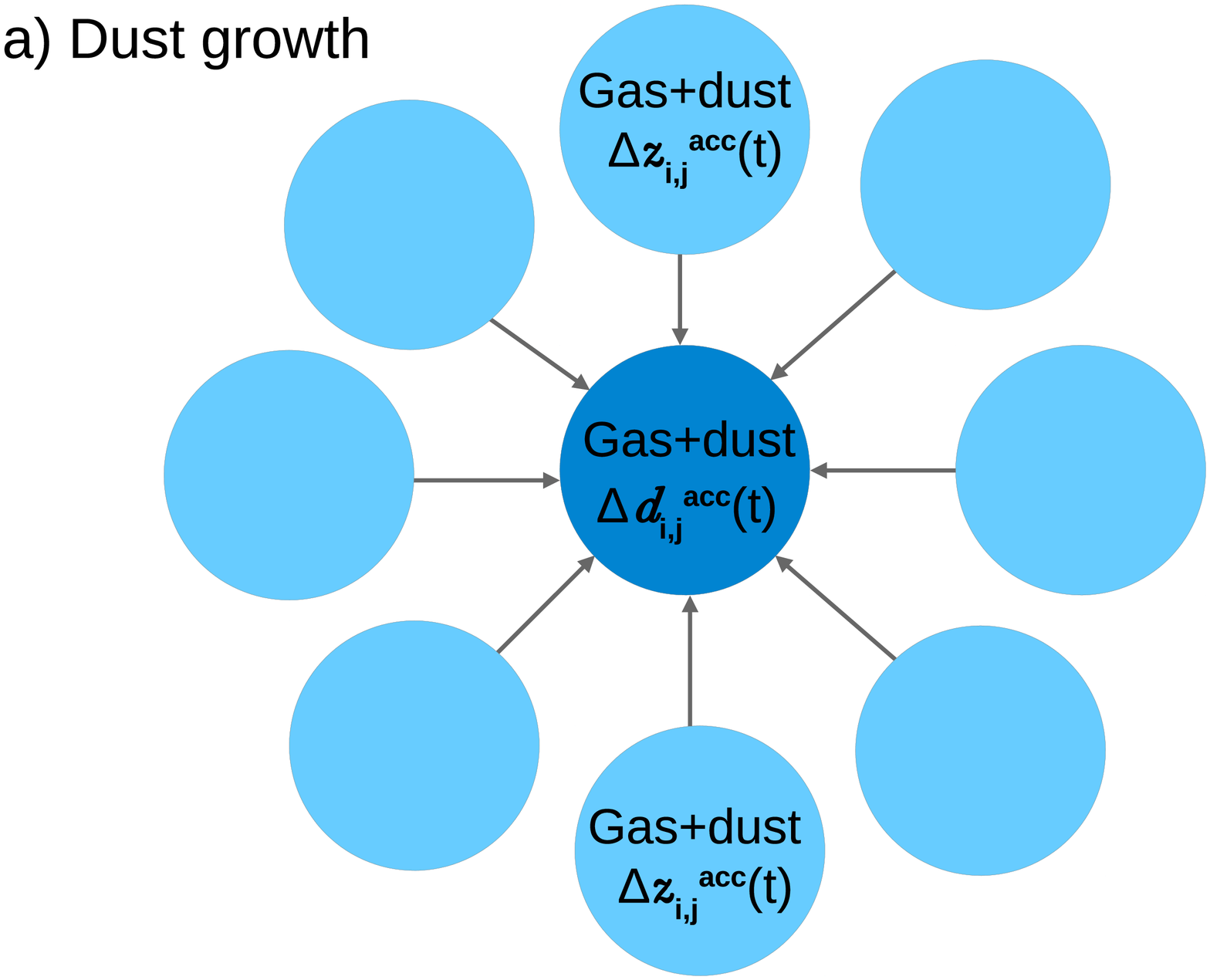}\par 
     \includegraphics[width=1.\linewidth]{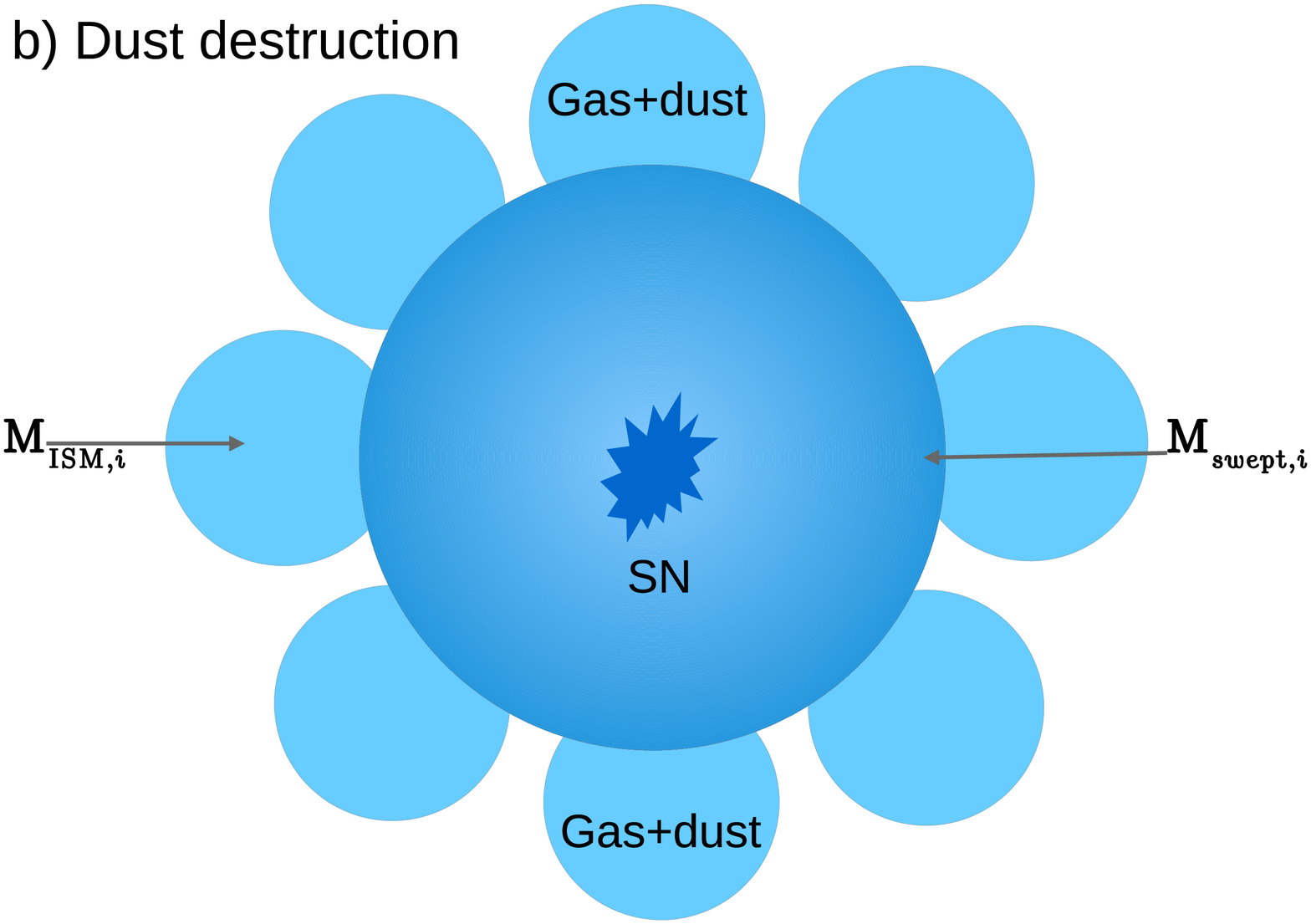}\par 
\end{multicols}
\caption{A schematic illustration of how dust growth and dust destruction are implemented in the simulation. Left: the dust growth process, sky blue and dark blue circles represent SPH gas particles. A fraction of the gas--phase metals ($\Delta z_{m,j}^{\rm acc}(t)$) is accreted onto dust grains in the central dark blue circle from all the surrounding SPH particles within the smoothing length. The total increase in dust mass in the central SPH is then $\Delta d_{i,j}^{\rm acc}(t)= \sum_{m=1}^{N{\rm nei, \it i}}\Delta z_{m,j}^{\rm acc}(t)$. Right: dust destruction process, sky blue is the same as in the left, while the central darker blue circle represents an expanding SN bubble. The bubble sweeps up the dust and gas in the surrounding SPH particles as it expands (the overlap region between the central and surrounding circles). In the process, part of the swept up dust is destroyed. M$_{ISM,i}$ and M$_{sweept,i}$ are the mass of the gas particle and the mass swept up by the SN, respectively.}
\vspace{-0.3cm}
\label{fig9}
\end{figure*}


\section{The model}
In this section, we present a modified version of the chemodynamical model presented in Bekki (2013; hereafter B13). In the present version, dust destruction and growth processes are modelled in such a way that they depend on the underlying ISM properties such as density, temperature, and metallicity. The rest of the model aspects remain the same. Hence the present version is more suitable for our purpose since it allows variations of the dust abundance between individual gas particles. This modification, further, allows the comparison with the spatially resolved 2D maps of observed galaxies. In the following, we briefly describe the model (equations and expressions presented here are the same as in B13 with a few exceptions) and the reader is referred to B13 for more details.
\subsection{Chemical enrichment}
The time evolution of eleven chemical elements such as He, C, N, O, Mg and Ca, ejected by SNIa, SNII, and AGB stars is followed. SNIa, SNII, and AGB stars are also dust formation sites where some of the produced elements are incorporated in dust grains or accreted on dust grains surfaces in the ISM. Thus the time evolution of those elements allows prediction of dust abundance and properties.

The amount of an element $j$ ($j$ = 1--11) ejected by $k$th stellar particle with metallicity $Z_k$ is assumed to be equally distributed across $N_{nei,k}$ neighbour SPH gas particles surrounding the $k$th stellar particle. Accordingly, the mass increase of each $j$ chemical element in $i$th SPH particle at time t is given by the following equation: 
\begin{equation}
\Delta z_{i,j}^{\rm ej}(t) = \sum_{k=1}^{N_{\rm nei, \it i}} 
m_{\rm s, \it k} Y_{k,j}(t-t_k)/N_{\rm nei, \it k},
\end{equation}
where $m_{\rm s, \it k}$ is the mass of $k$th stellar particle, $Y_{k,j}(t-t_k)$
is the mass of each $j$ chemical element ejected from stars per unit mass at 
time $t$, and $t_k$ represents the time when the $k$th stellar particle is 
born from a gas particle. A non--instantaneous recycling of those elements is accounted for.

\subsection{Dust model}
We adopt the dust model proposed by D98, and it consists of the following processes I--IV which are described in the following for the current model.

(I) Dust formation in stellar winds of  AGB stars, SNIa, and SNII. The total mass of $j$th element ($j$ = C, O, Mg, Si, S, Ca, and Fe) in dust from $k$th type of stars (SNIa, SNII, and AGB stars) is given by the following equation:
\begin{equation}
m_{\rm dust, \it j}^k= \delta_{\rm c, \it j}^k F_{\rm ej}(m_{\rm ej, \it j}^k),
\end{equation}
where $\delta_{\rm c, \it, j}^k$ is the condensation efficiency for each $j$ chemical element from the $k$th stellar type, $F_{\rm ej}$ is a function that determines the amount  of metals that is usable for dust formation, and $m_{\rm ej, j}^k$ is the mass of the $j$th element ejected from the $k$th stellar type. The stellar yield tables by Tsujimoto et al. (1995) for SNIa and SNII, and van den Hoek \& Groenewegen (1997) for AGB stars are used to estimate the total mass of stellar ejecta.  Details of dust yields, $\delta_{\rm c, \it j}^k$, and $F_{\rm ej}(m_{\rm ej, \it j}^k)$ are given in B13. 
\begin{table}
\centering
\begin{minipage}{75mm}
\caption{Description of the basic parameter values
for the MW--like model.}
\begin{tabular}{cc}
\hline
{Physical properties}
& {Parameter values}\\
\hline
{Total Mass \footnote{$M_{\rm h}=M_{\rm dm}+M_{\rm g}$, where
$M_{\rm dm}$ and $M_{\rm g}$ are the total masses of dark matter halo
and gas in a galaxy, respectively.}}
 & $M_{\rm h}=10^{12} {\rm M}_{\odot}$  \\
{Structure \footnote{ For the structure of the dark matter halo NFW profile with a virial radius ($r_{\rm vir}$) and a $c$--parameter is adopted.}}
 & $r_{\rm vir}=245$ kpc,  $c=10$  \\
Gas fraction & $f_{\rm g}=0.1$     \\
Initial H$_2$ fraction & 0.01     \\
Initial metallicity/gradient   &   ${\rm [Fe/H]_0}= 0.30$ dex/$-0.04$ dex/kpc \\
Initial (dust/metal) ratio  & 0.4  \\
{SF \footnote{$\rho_{\rm th}$ is the threshold gas density for star formation. The interstellar radiation field (ISRF) is included in the estimation of  ${\rm H_2}$ mass fraction in this model.}}
 & ${\rm H}$--dependent,  ISRF,  $\rho_{\rm th}=1$ cm$^{-3}$ \\
IMF & Salpeter ($\alpha=2.35$) \\
Softening length  & $\varepsilon_{\rm dm}=935$ pc, $\varepsilon_{\rm g}=94$ pc \\
Gas mass resolution   & $m_{\rm g}=3\times10^4 {\rm M}_{\odot}$ \\
\hline
\end{tabular}
\end{minipage}
\end{table}

(II) Accretion of the ISM gas--phase metals on dust grains surfaces. This process is mainly characterized by the accretion timescale ($\tau_{\rm acc}$). Since in this study $\tau_{\rm acc}$ is different for different gas particles depending on their properties, $\tau_{\rm acc}$ is indicated by $\tau_{\rm acc, \it i}$ for the $i$th gas particle. Hence the mass increase of the $j$th element ($j$ = C, O, Mg, Si, S, Ca, and Fe) in dust due to accretion is given by the following equation:
\begin{equation}
\Delta d_{i,j}^{\rm acc}(t)=\Delta t_i (1-f_{\rm dust,\it i, j}) 
d_{i,j}(t) /\tau_{\rm acc, \it i},
\end{equation} 
where $\Delta t_i$ is the individual time step width for $i$th gas particle and $f_{\rm dust, \it i, j}$ is the fraction of the $j$th chemical element that is locked up in dust (see also subfigure a in Fig. \ref{fig9}). We adopt the formula presented by Asano et al. (2013; Eq. 20) for $\tau_{\rm acc, \it i}$.

\begin{equation}
\begin{split}
C_s \tau_{\rm acc, \it i} =  2\times 10^7 \times (\frac{\bar{a}}{0.1\micron}) \times (\frac{n_{H,i}}{100cm^{-3}})^{-1}\\ \times (\frac{T_{g,i}}{50K})^{-\frac{1}{2}} \times (\frac{Z_i}{0.02})^{-1} yr,
\end{split}
\end{equation}
where $C_s$, $\bar{a}$, $n_{H,i}$, $T_{g,i}$, and $Z_i$ are the dust sticking coefficient, typical size of grains (taken to be $0.1\micron$), hydrogen number density, temperature, and metallicity of the $i$th gas particle, respectively. Dust growth is switched on in all gas particles, i.e. it occurs throughout the ISM, however, with a rather long timescale in low--density environments and vice versa. This process leads to depletion of gas--phase elements, accordingly, the mass of the $j$th element in the gas decreases by the same amount, i.e. 

\begin{equation}
\Delta z_{i,j}^{\rm acc}(t)= - \Delta d_{i,j}^{\rm acc}(t).
\end{equation}

(III) Dust destruction by energetic SNe explosions. This process decreases the mass of $j$th element in dust and increases its amount in the gas by the same value. Here we do not use the destruction timescale as is usually used in literature to characterize dust destruction (e.g. Dwek \& Scale 1980; Mckee 1989; D98; B13), we implement the process according to the following steps instead:

1. The location of SNII events is identified at each time step in the simulation. All SNe deposit the same amount of energy into the ISM (see section 2.5).

2. Gas particles within the smoothing length ($\varepsilon_{\rm g}$) from the centre of the $k$th SN are identified as neighbours ($N_{\rm nei, k}$). The number of neighbours is approximately 50.

3. Estimating the destruction efficiency. First, the density and mass of the ISM ($n_{H,k}$ and $M_{ISM,k}$) surrounding the $k$th SN are estimated by averaging over the density and mass of the gas in its neighbour SPH particles as defined in step 2. Then, we use Eq. A3 and the overall results for model C20 in Table 8 in Nozawa et al. 2006 to calculate the destruction efficiency $F_{des,k}$.

\begin{equation}
 log F_{des, k} = a(log n_{H, k})^2 + b(log n_{H,k}) + c,
\end{equation}
where $a = -0.067$, $b = 0.33$, and $c = -0.47$. Nozawa et al. 2006 also defines the destruction efficiency as the ratio between the mass of dust destroyed and the mass of dust swept up by the shock. $F_{des, k}$ is then normalized by 0.34 ($F_{des, k}$ for n$_H$ = 1 $cm^{-3}$). Additionally, we assume that dust is well mixed in the gas and hence the fraction of the ISM swept up by SN ($\frac{M_{swept,k}}{M_{ISM,k}}$) equals to the fraction of dust swept up ($\frac{M_{dust, swept,k}}{M_{dust,k}}$). M$_{swept,k}$ is defined as the mass of gas swept up by SN shock until the shock velocity decelerates below 100 km/s. Because M$_{ISM,k}$ is approximately the same for all SNe (the mass and number of neighbour gas particles is approximately constant), the swept up dust fraction is proportional to M$_{swept,k}$ ($M_{swept,k} \propto n_{H,k}^{-0.142}$; Nozawa et al. 2006 Eq. A4). This assumption together with the destruction efficiency definition from Nozawa et al. allow us to estimate the total mass of dust destroyed. We indicate this quantity by $\epsilon_{des,k}$.

4. The final step involves sharing the total mass of dust destroyed equally among the $N_{\rm nei,k}$ gas particles, i.e. reducing the dust mass in each $i$ gas particle by a certain amount, $\epsilon_{des,i}$.

\begin{equation}
\epsilon_{des,i} = f_{des}\epsilon_{des,k},
\end{equation}
where $f_{des}$ is a parameter that determines the fraction of dust destroyed in a single SPH particle surrounding SN.  Accordingly, the total amount of dust destroyed by the $k$th SN in our models ($\epsilon_{des,i}\times N_{\rm nei,k}$) is no longer the same as Nozawa's unless $f_{des}$ equals 0.02 and $N_{\rm nei,k}$ equals 50. Fig. \ref{fig9}b illustrates the destruction process in our models. The change in the mass of the $j$th element in dust due to destruction is hence

\begin{equation}
\Delta d_{i,j}^{\rm des}(t)= - \epsilon_{des,i}.
\label{e8}
\end{equation}

In the previous version of the code (B13), the change in the mass of the $j$th element in dust due to destruction is estimated using the following equation:

\begin{equation}
\Delta d_{i,j}^{\rm dest}(t)= -\Delta t_i 
d_{i,j}(t) /\tau_{\rm dest, \it i},
\end{equation}
where $\Delta t_i$ and  $\tau_{\rm dest, \it i}$ are the individual time step width and destruction timescale for the $i$th gas particle ($\tau_{\rm dest, \it i}$ is taken to be constant). The present implementation (Eq. \ref{e8}) improves this considerably.

It is worthwhile mentioning that, the low metallicity homogenous ISM in high redshift galaxies for which  Nozawa et al. did their calculations is different from the ISM in low redshift galaxies. The ISM of low redshift galaxies is richer in metals and is inhomogenous with multiple phases that have different densities and filling factors. Thus, SNe shocks predominantly propagate in the diffuse low--density regions and they become inefficient if the density contrast between different phases is high (Dwek et al. 2007). The presence of abundant metals also reduces the ability of the shock to propagate since they provide efficient cooling (M$_{swept,k}$ is reduced; Yamasawa et al. 2011; Asano et al. 2013).  Furthermore, massive stars that migrate away from their parent clusters and explode in diffuse media have higher destruction effect than those exploding in high--density media (e.g. McKee 1989; Oey \& Lamb 2012) or as part of correlated SNe (exploding in superbubbles of previous SNe; Higdon \& Lingenfelter 2005). Not accounting for or resolving those different effects results in estimation errors of the dust destruction that go both ways (i.e. underestimation and overestimation).

(IV) polycyclic aromatic hydrocarbons (PAHs) formation. The most likely site for PAH formation is the carbon--rich ABG stars (Draine et al. 2007; Meixner et al. 2010; Takagi et al. 2010; Sandstrom et al. 2012). Thus, the assumption is that some fraction of the carbon dust produced by carbon--rich AGB stars can eventually become PAH dust. Properties of PAH dust is outside the scope of this study, refer to B13 for more details.

\begin{table}
\begin{minipage}{75mm}
\caption{Description of the $\chi^2$ fit parameter symbols used in this study for the dust spatial correlations.}
\begin{tabular}{cc}
\hline 
{Symbol}
& {The meaning}\\
\hline
\multirow{1}{*}{$\alpha_{H_{2}l}$}  &slope of the linear correlation \\& between $\Sigma_{H_{2}}$ and $\Sigma_D$\\
\multirow{1}{*}{$\alpha_{Gl}$} & slope of the linear correlation  \\& between $\Sigma_{G}$ and $\Sigma_D$\\
\multirow{1}{*}{$\alpha_{H_{2}nl}$} & slope of the nonlinear correlation \\ & between $\Sigma_{H_{2}}$ and $\Sigma_D$\\
\multirow{1}{*}{$\alpha_{Gnl}$} & slope of the nonlinear correlation  \\ & between $\Sigma_{G}$ and $\Sigma_D$\\
\multirow{1}{*}{P} & the power index in the nonlinear correlation\\ & case\\
\hline
\end{tabular}
\end{minipage}
\label{table2}
\end{table} 
\subsection{H$_2$ Formation and Dissociation}
The balance between H$_2$ formation on dust grains and dissociation by FUV  radiation is what determines the mass fraction of H$_2$ to the total hydrogen gas ($f_{\rm H_2}$) in B13 and the present model. $f_{\rm H_2}$ is a function of the ISM properties and the interstellar radiation field (ISRF). 
\begin{equation}
f_{\rm H_2} = F(T_{g,i}, n_{H,i}, D_i, \chi_i),
\end{equation}
where F is a function for $f_{\rm H_2}$ determination and $T_{g,i}$, $n_{H,i}$, $D_i$, and  $\chi_i$ are the $i$th gas particle temperature, hydrogen number density, dust--to--gas ratio, and the strength of the FUV radiation field around the gas particle, respectively, calculated at each time step. To estimate $\chi_i$ SEDs of stellar particles around each $i$ gas particle (thus ISRF) are first estimated for ages and metallicities of the stars using stellar population synthesis codes for a given IMF (initial mass function; Bruzual \& Charlot 2003). Then the strength of the FUV part of the ISRF is estimated from the SEDs so that $\chi_i$ can be derived for the $i$th gas particle. Salpeter IMF is adopted in this model.

\begin{table*}
\centering
\caption{The grid of models with different combinations of $f_{des}$ and $C_s$ values. $\beta_{sm}$ is the slope of the linear fit to the simulated data in $log(D/G)$ vs $12+log(\frac{O}{H})$ relation.  $M_{H_2}$, $M_{HI}$, $M_{\ast}$, $M_{D}$, $M_{Z}$, and $M_{G}$ are the total mass of H$_2$, HI, stars, dust, metals, and total gas. $\beta_{sm}$, $M_{H_2}$, $M_{HI}$, $M_{\ast}$, $M_{D}$, $M_{Z}$, and $M_{G}$ are measured at T = 1 Gyr.}

\begin{tabular}{llllllllllllll}
\hline
{Model ID}  
 &  {$f_{des}$}
 &  {$C_s$}
 & {$\beta_{sm}$} 
 &  {$\frac{M_{H_2}}{M_{HI}}$}
 & {$\frac{M_{H_2}}{M_{\ast}}$}
 &  {$\frac{M_{D}}{M_{Z}}$}
 & {$\frac{M_{D}}{M_{\ast}}$} 
 & {$\frac{M_{Z}}{M_{\ast}}$} 
 & {$\frac{M_{D}}{M_{G}}$}\\
\hline
M1 &  0.01 & 1 & 1.41$\pm$0.001 & 0.484 & 0.029 & 0.735 & 0.0009 & 0.0012 & 0.011 \\
M2 &  0.02 & 1 & 0.59$\pm$0.001 & 0.204 & 0.015 & 0.291 & 0.0005 & 0.0017 & 0.006 \\
M3 &  0.03 & 1 & $-0.46\pm$0.001 & 0.118 & 0.009 & 0.147 & 0.0003 & 0.0019 & 0.003 \\
M4 &  0.05 & 1 & $-2.16\pm$0.001 & 0.063 & 0.005 & 0.077 & 0.0002 & 0.0021 & 0.002 \\
M5 &  0.01 & 0 & $-0.60\pm$0.001 & 0.082 & 0.007 & 0.127 & 0.0003 & 0.0020 & 0.003 \\
M6 &  0.01 & 0.1 & $-0.33\pm$0.001 & 0.084 & 0.008 & 0.144 & 0.0003 & 0.0020 & 0.003 \\
M7 &  0.01 & 0.2 & $-0.07\pm$0.001 & 0.10 & 0.008 & 0.167 & 0.0003 & 0.0019 & 0.004 \\
M8 &  0.01 & 0.3 & 0.25$\pm$0.001 & 0.119 & 0.009 & 0.203 & 0.0004 & 0.0019 & 0.004 \\
M9 &  0.01 & 0.4 & 0.53$\pm$0.001 & 0.157 & 0.012 & 0.250 & 0.0004 & 0.0018 & 0.005 \\
M10 &  0.01 & 0.5 & 0.73$\pm$0.001 & 0.20 & 0.015 & 0.30 & 0.0005 & 0.0017 & 0.006 \\
M11 &  0.01 & 0.6 & 0.95$\pm$0.001 & 0.253 & 0.018 & 0.372 & 0.0006 & 0.0016 & 0.007 \\
M12 &  0.01 & 0.7 & 1.13$\pm$0.001 & 0.732 & 0.037 & 0.469 & 0.0007 & 0.0015 & 0.008 \\
M13 &  0.01 & 0.8 & 1.22$\pm$0.001 & 0.346 & 0.022 & 0.535 & 0.0008 & 0.0015 & 0.010 \\
M14 &  0.01 & 0.9 & 1.35$\pm$0.001 & 0.40 & 0.025 & 0.644 & 0.0009 & 0.0014 & 0.010 \\
M15 &  0.02 & 0 & $-1.52\pm$0.001 & 0.056 & 0.005 & 0.084 & 0.0002 & 0.0021 & 0.002\\
M16 &  0.02 & 0.1 & $-1.36\pm$0.001 & 0.061 & 0.005 & 0.091 & 0.0002 & 0.0021 & 0.002 \\
M17 & 0.02 &  0.3 & $-0.97\pm$0.001 & 0.080 & 0.006 & 0.109 & 0.0002 & 0.0020 & 0.003 \\
M18 &  0.02 & 0.5 & $-0.50\pm$0.001 & 0.090 & 0.007 & 0.138 & 0.0003 & 0.0020 & 0.003 \\
M19 &  0.02 & 0.7 & $-0.01\pm$0.001 & 0.114 & 0.009 & 0.183 & 0.0003 & 0.0019 & 0.004 \\
M20 &  0.02 & 0.8 & 0.23$\pm$0.001 & 0.137 & 0.011 & 0.216 & 0.0004 & 0.0018 & 0.005 \\
M21 &  0.02 & 0.9 & 0.45$\pm$0.001 & 0.160 & 0.012 & 0.260 & 0.0005 & 0.0018 & 0.005 \\
\hline
\end{tabular}
\vspace{-0.15cm}
\label{table3}
\end{table*} 
 
\subsection{Star formation}
Recent observations showed that local SFRs are well correlated with local dust density rather than with the atomic hydrogen (HI) or H$_2$ densities (Komugi et al. 2018). However, in this model, we implement the commonly used `H--dependent' SF recipe, i.e. SFR depends on the total gas density.  In this recipe, if a gas particle satisfies the following conditions (i)--(iii), a new stellar particle is born out of the gas with Salpeter IMF. (i) The sound crossing timescale is longer than the local dynamical timescale, (ii) the local velocity field is identified as being consistent with gravitationally collapsing, and (iii) the local density exceeds a threshold density for star formation ($\rho_{\rm th}$). In the present study $\rho_{\rm th}$ is taken to be 1 $cm^{-3}$.

\subsection{Gravitational dynamics and hydrodynamics}
Multiple gravitational softening lengths are used in this model, i.e. $\varepsilon_{\rm dm}$ for the dark matter component and $\varepsilon_{\rm g}$ for the gas component. Furthermore, new stars have the same softening length as the gas particles. In case two different components interact, the mean of their softening lengths is taken. The ISM is considered to be an ideal gas with a ratio of specific heats equals 5/3, and Hernquist \& Katz (1989) method is used to implement the SPH.  Moreover, the predictor--corrector algorithm is adopted to integrate the equations describing the time evolution of the system. An individual time step width ($\Delta t$) is allocated for each particle that is determined by its physical properties. The maximum time step width ($\Delta t_{\rm max}$) is $0.01$ in simulation units which means   $\Delta t_{\rm max}=1.41 \times 10^6$ yr in physical units and  $\Delta t_{\rm min}=$1.41 $\times 10^4$ yr.

Each SN is assumed to eject energy ($E_{\rm sn}$) of $10^{51}$ erg, 90\%  of $E_{\rm sn}$ is used for increasing the thermal energy (`thermal feedback'), and the rest is exhausted by random motions (`kinematic feedback'). The energy--ratio of thermal to kinematic feedback is consistent with previous numerical simulations by Thornton et al. (1998). The way to distribute $E_{\rm sn}$ of SNe among neighbour gas particles is the same as described in Bekki et al. (2013). The radiative cooling processes are included by using the cooling curve by Rosen \& Bregman (1995) for  $100 \le T < 10^4$K and the MAPPING III code for $T \ge 10^4$K (Sutherland \& Dopita 1993). 

\subsection{Isolated disc galaxy model}
In the present model, the Navarro, Frenk \& White (1996) dark matter halo density profile is adopted.
\begin{equation}
{\rho}(r)=\frac{\rho_{0}}{(r/r_{\rm s})(1+r/r_{\rm s})^2},
\end{equation}
where  $r$, $\rho_{0}$, and $r_{\rm s}$ are the spherical radius, the characteristic density, and the scale length of the halo, respectively. The $c$--parameter ($r_{\rm vir}/r_{\rm s}$, where $r_{\rm vir}$ is the virial radius) is chosen for a given $M_{\rm h}$ (= M$_{dm}$ + M$_{g}$ where  M$_{dm}$ and M$_{g}$ are the mass of the dark matter and gaseous halos, respectively) according to the predicted $c$--$M_{\rm h}$ relation in the $\Lambda$CDM simulations (e.g. Neto et al. 2007). The gaseous halo is assumed to have the same density distribution initially and to be in hydrostatic equilibrium. Therefore, the initial temperature of a halo gas particle is determined by the gravitational potential at its location, its total mass, and density via Euler's equation for hydrostatic equilibrium.
\begin{figure*}
\begin{multicols}{2}
    \includegraphics[width=1.\linewidth]{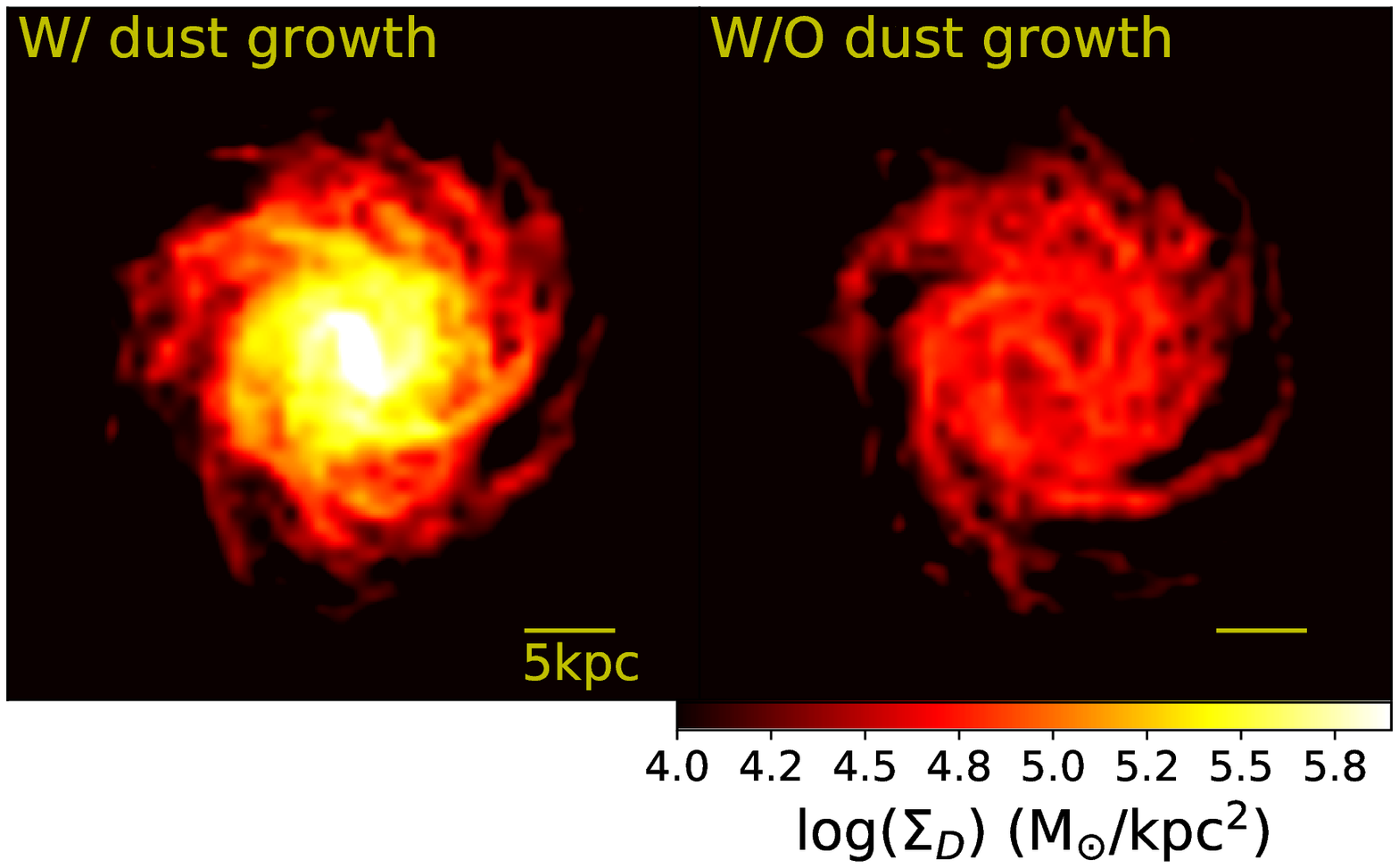}\par 
     \includegraphics[width=1.\linewidth]{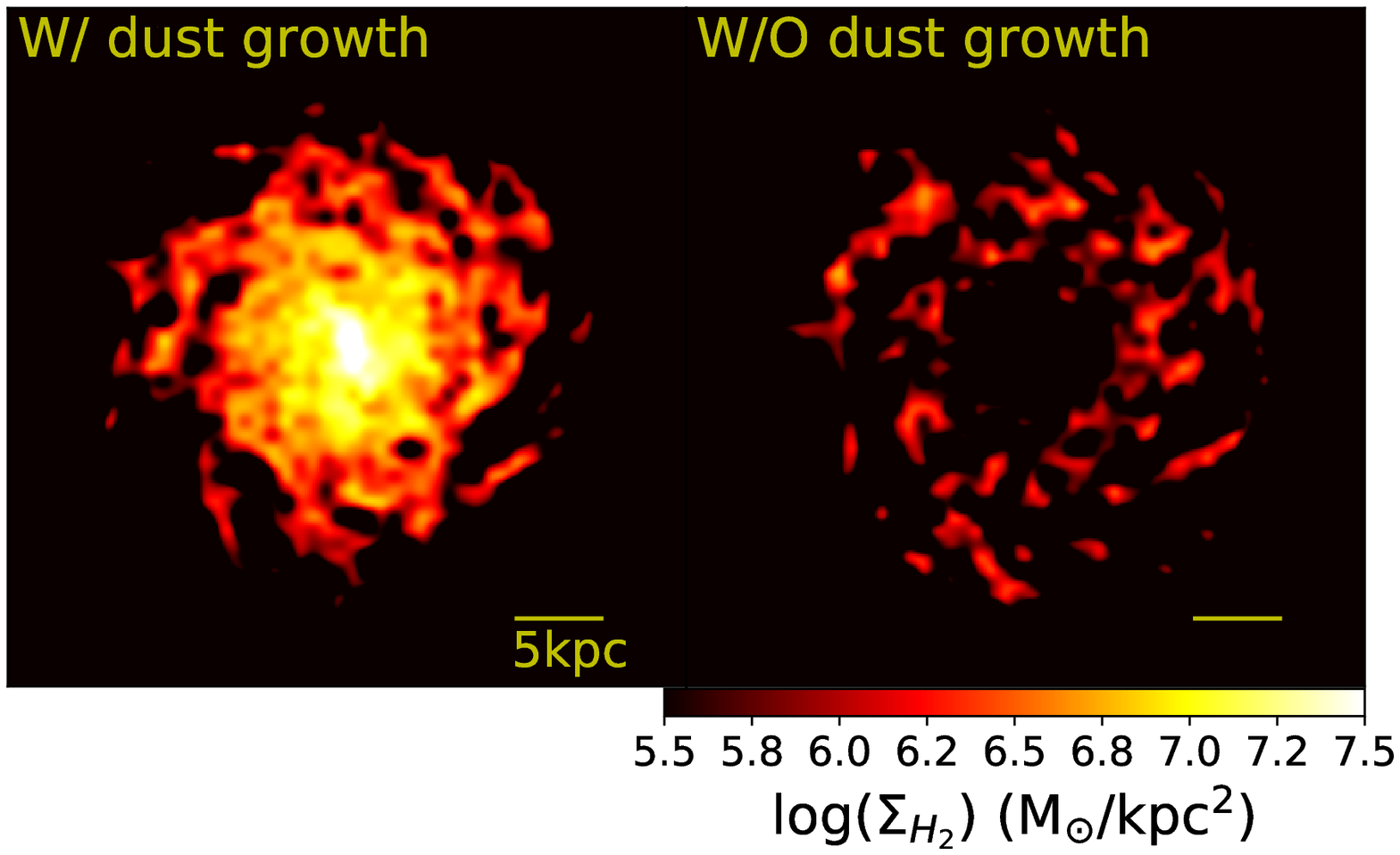}\par 

    \end{multicols}
\begin{multicols}{2}
    \includegraphics[width=1.\linewidth]{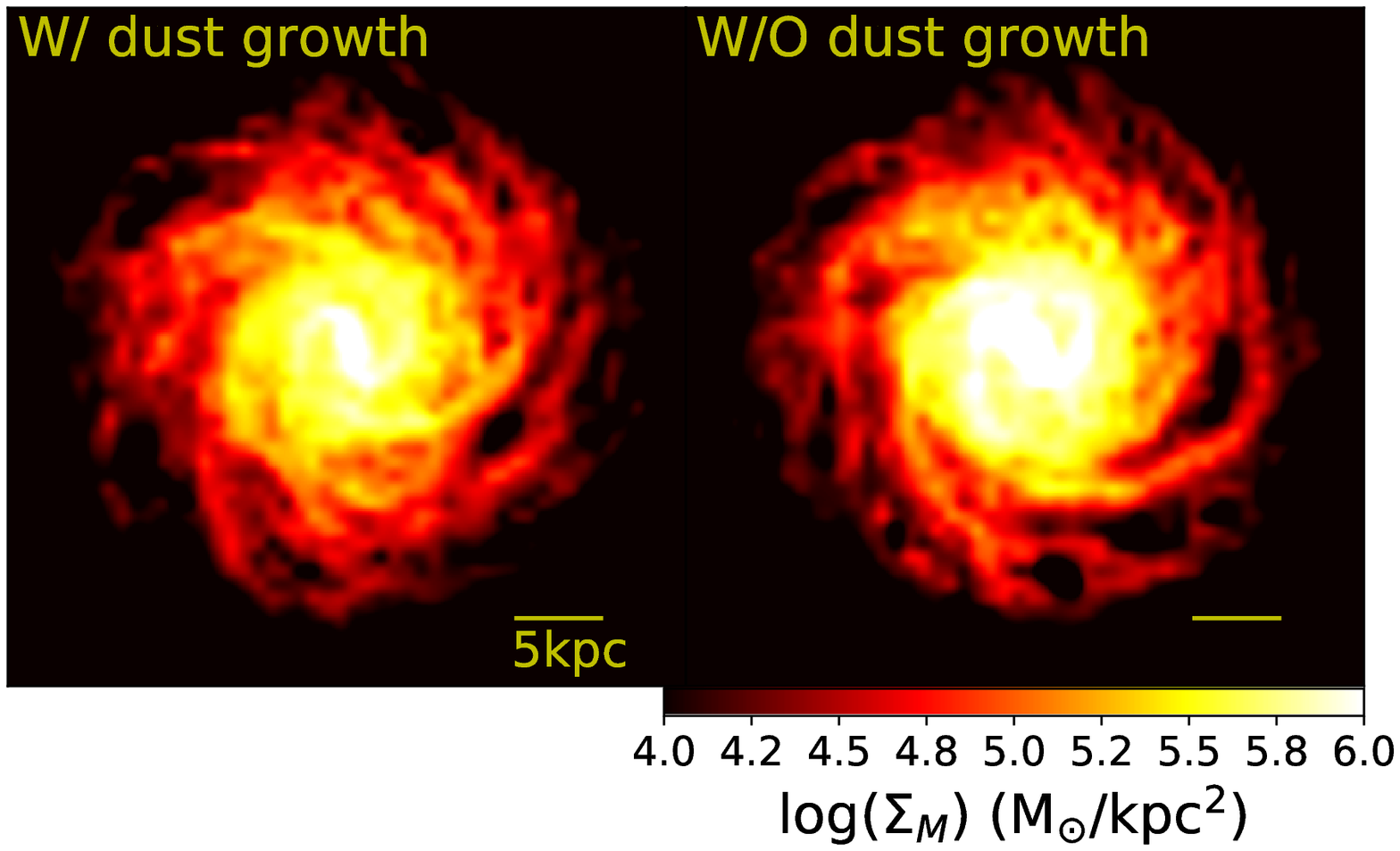}\par 
    \includegraphics[width=1.\linewidth]{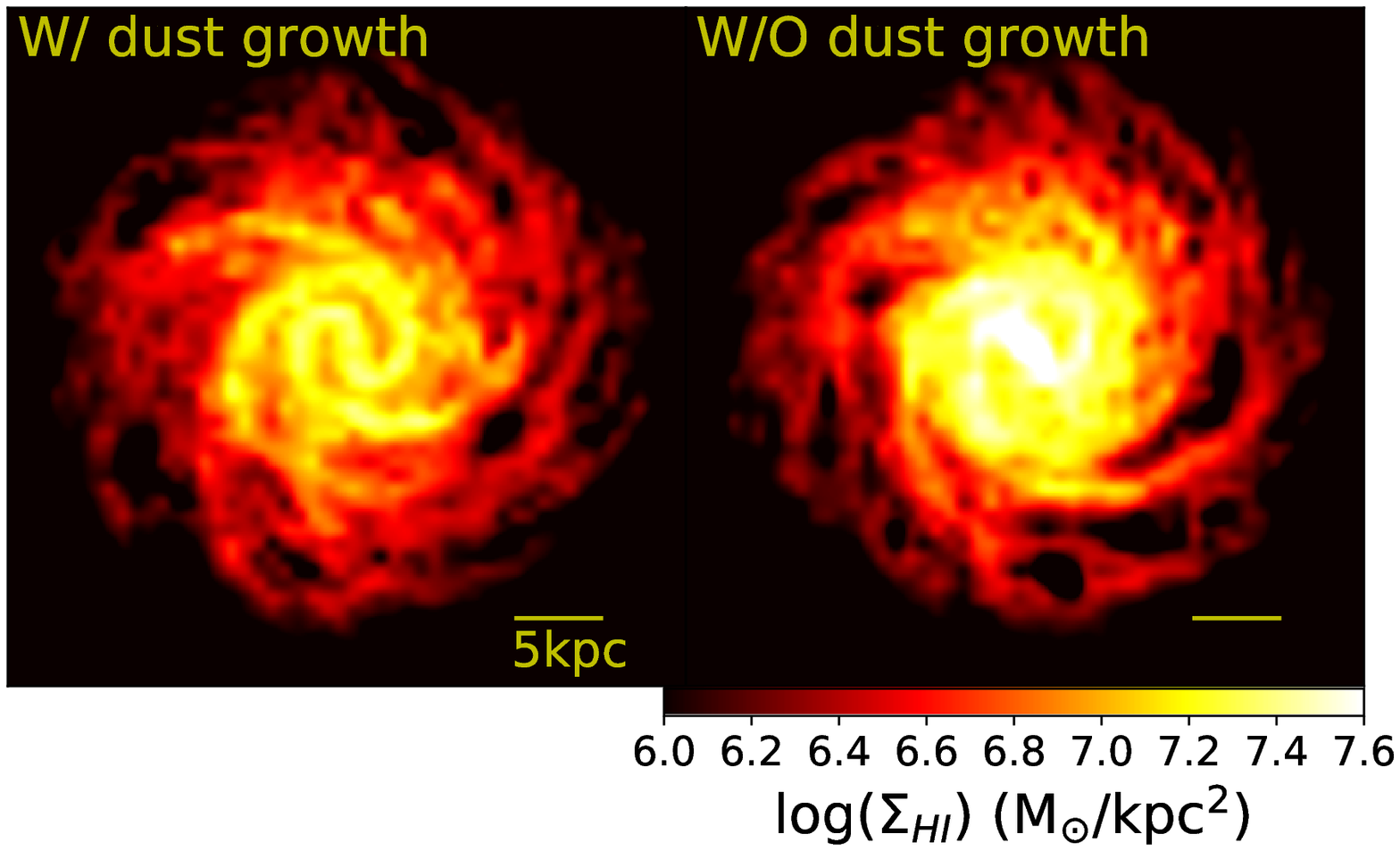}\par 
    \end{multicols}
\caption{The surface mass density of dust (upper left), H$_2$ (upper right), metals (bottom left), and  HI (bottom right) ($\Sigma_D$, $\Sigma_{H_2}$,  $\Sigma_M$, and $\Sigma_{HI}$, respectively) at T = 1 Gyr  in logarithmic scale projected onto $xy$ plane for the MW models, M1 (left panel of each subfigure, $f_{des}$ = 0.01, $C_s$ = 1) and M5 (right panel of each subfigure, $f_{des}$ = 0.01, $C_s$ = 0).}
\vspace{-0.3cm}
\label{fig1}
\end{figure*}
\subsection{$f_{des}$ and $C_s$ investigated ranges}
In our framework, $f_{des}$ and $C_s$ are the two free parameters that modify the dust growth rate and the destroyed dust fraction (see section 2.2). Thus, we focus on studying those two parameters by investigating a gird of 21 models with different combinations of $f_{des}$ and $C_s$ values ranging between 0.01 and 0.05 for $f_{des}$, and between 0 and 1 for $C_s$. It is important to stress that, although this study is based on MW--like galaxy models with the typical MW properties listed in Table 2, they are not calibrated to reproduce the global values such as dust--to--gas, dust--to--metals ratios. These properties are the result of the dust evolution. To access the feasibility of the models, we use the $log(D/G)$ vs $[12+log(\frac{O}{H})]$  scaling relation. This relation is essential in tracking dust evolution in galaxies and its processing in the ISM.  However, it is not clear whether or not the slope of the relation is unity.  In the Milky way, D98 found that the slope of the $log(D/G)$ vs $[12+log(\frac{O}{H})]$ relation is unity assuming the same time dependence for the destruction and growth timescales. Remy--Ruyer et al. 2014 also showed that in galaxies with $[12+log(\frac{O}{H})]$ $>$ 7.96 $\pm$ 0.47 or 8.10 $\pm$ 0.43,  a unity power index is a good fit to the relation although the scatter is large. Remy--Ruyer et al. fit could describe, on average, the integrated dust abundance in those galaxies as a function of the global metallicity. In dwarf (Remy--Ruyer et al. 2014) and resolved spiral galaxies (e.g. M101 galaxy; Chiang et al. 2018), a unity slope does not accurately explain the relation between dust and metallicity. Accordingly, we only require the dust--to--gas ratio to increase monotonically as a function of metallicity.

\begin{figure*}
\begin{multicols}{2}
     \includegraphics[width=1.\linewidth]{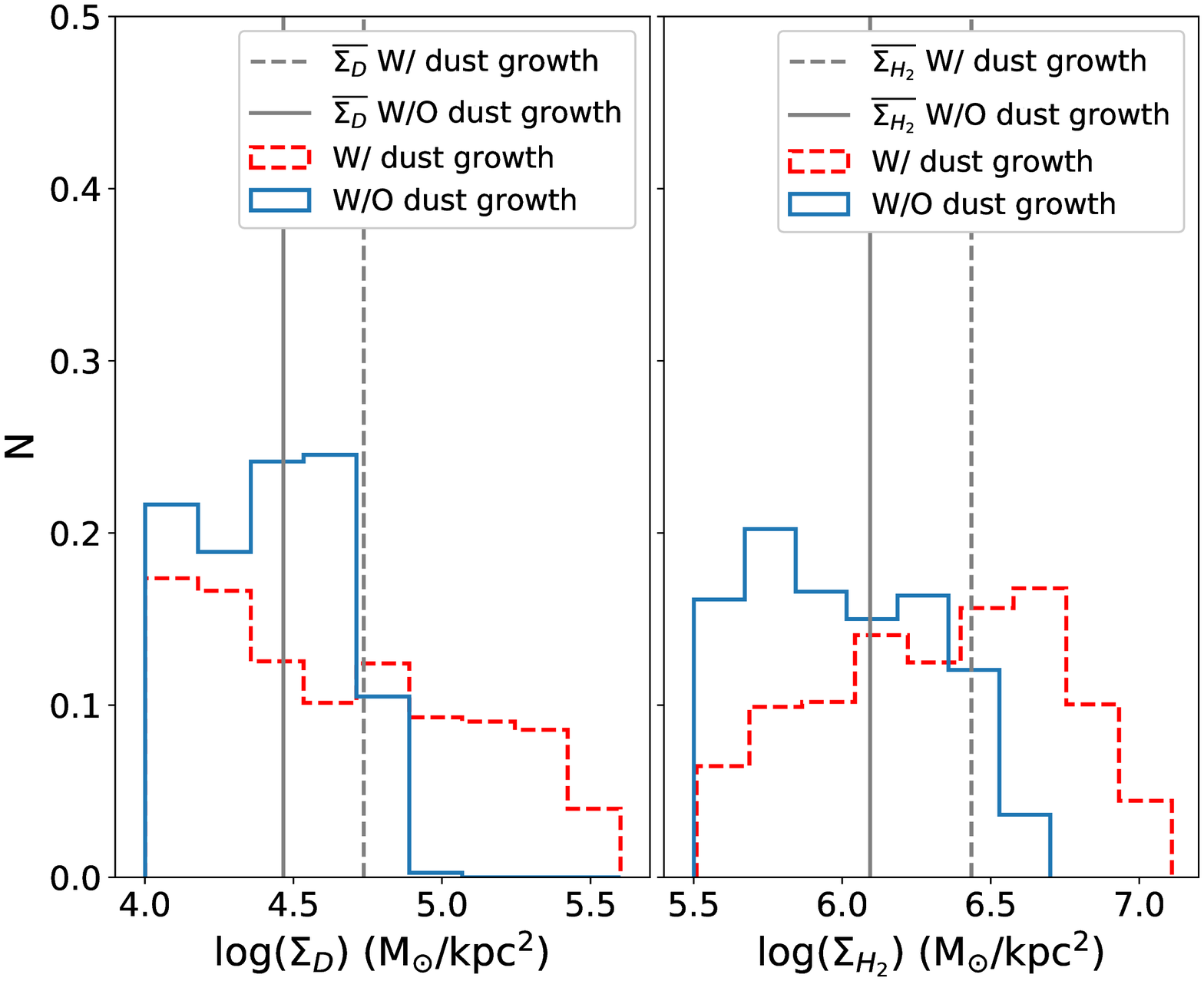}\par 
    \includegraphics[width=1.\linewidth]{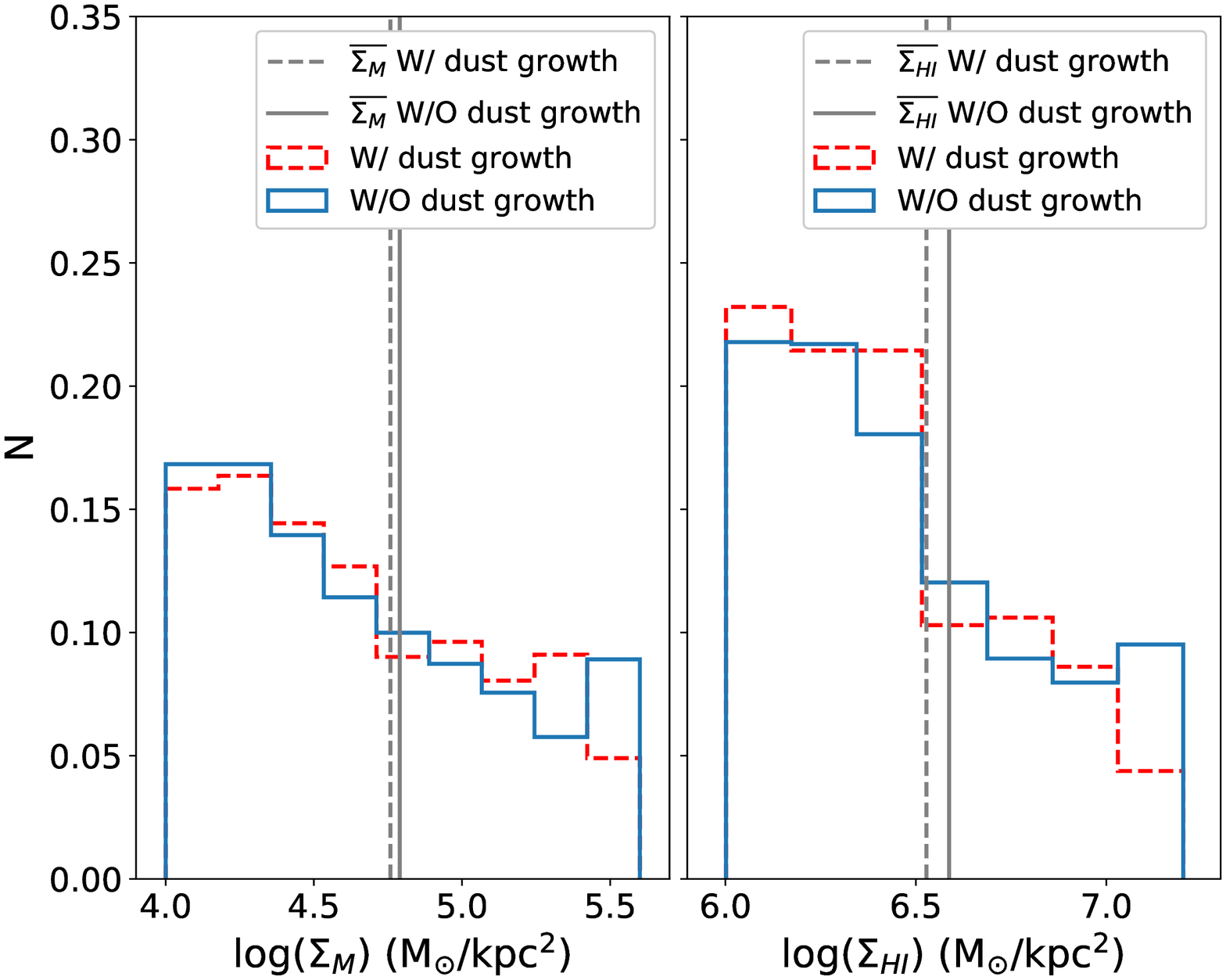}\par 
    \end{multicols}
\caption{Normalised histograms of the surface mass density of the dust (left), H$_2$ (middle left), metals (middle right), and HI (right) ($\Sigma_D$, $\Sigma_{H_2}$,  $\Sigma_M$, and $\Sigma_{HI}$, respectively) in logarithmic scale at T = 1 Gyr for M1 (red dashed) and M5 (blue solid) models. The vertical gray dashed, and solid lines represent the mean of each surface density for the models M1 and M5, respectively.}
\vspace{-0.3cm}
\label{hist1}
\end{figure*} 

\begin{figure}
\begin{center}
\includegraphics[height=6cm,width=8cm]{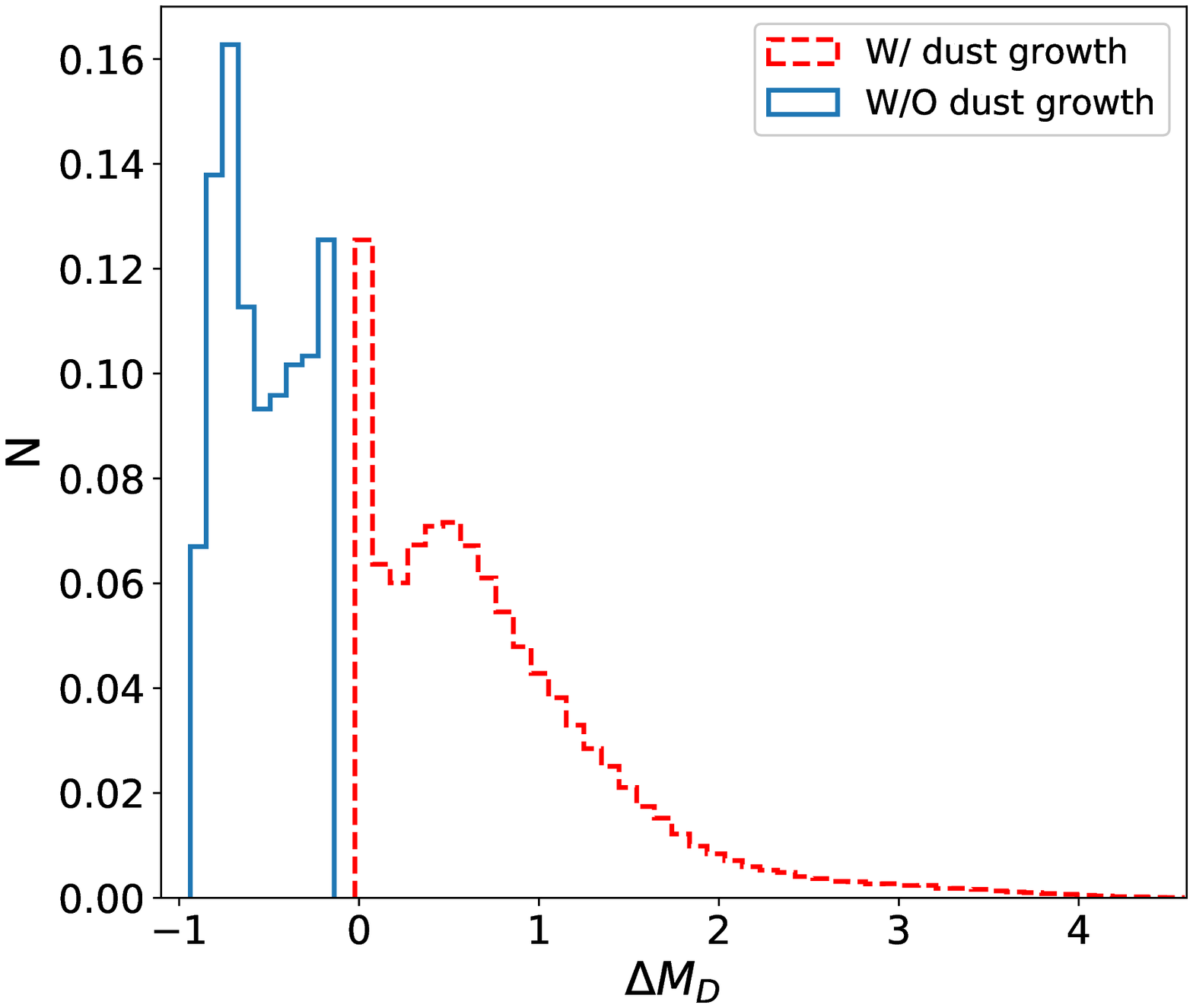}
\end{center}
\caption{Normalised histograms of the dust growth rate in 1 Gyr of evolution ($\Delta M_{D} = \frac{M_D(final) - M_D(initial)}{M_D(initial)}$) for the models M1 (red dashed) and M5 (blue solid).}
\label{hist3}
\end{figure}
 
Table 4 summarises the results of the $\chi^2$ fit to $log(D/G)$ = $\beta_{ms}$[$12+log(\frac{O}{H})] + b$ for each model at T = 1 Gyr. We first fix $C_s$ to its maximum value (one) and vary the value of $f_{des}$ for four different models (the first four rows in Table 4). We stop when a negative slope for the relation is obtained since such a trend suggests that as the galaxy evolves and increases its metals, dust mass decreases which is not physical and not observed. Accordingly, we discard all the models that behave in such a way, this process resulted in two values for $f_{des}$ 0.01 and 0.02. Fig. \ref{fig10} shows an example of two models (M5, M15) with negative slopes. Note that we have not studied the full parameter space for $f_{des}$ and stopped at $f_{des}$ = 0.05, and this is because of all models with $f_{des}$ $>$ 0.02 would yield negative slopes. 

The rest of the models are used to study variations of $C_s$ and their influence. For each value of $f_{des}$ obtained in the previous step,  we also vary the value of $C_s$ for different models, and again discard all models with negative slopes. As a final step, we use observational studies to clean the models' set farther by rejecting all models that yield slopes outside the observed range (below 0.55; Sandstrom et al. 2013). Several studies addressed this relation and tried to quantify it with different data sets on different scales. Therefore, we choose studies in such a way that they are statistically representative and on different scales, from integrated galaxy properties (Draine et al. 2007; Remy--Ruyer et al. 2014) to spatially resolved (Leroy et al. 2011; Sandstrom et al. 2013; Chiang et al. 2018). This resulted in $C_s$ values range between 0.5 and 1. However, applying a single value of $f_{des}$ and $C_s$ throughout a galaxy is an approximation (perhaps even oversimplification) of their complex behaviour.

The adopted SF recipe could influence the range of dust parameters. Implementing H$_2$--dependent recipe, for instance, would suppress star formation in the first Gyr of evolution. This is because, despite the high total density of the gas particle, H$_2$ forms a small fraction of the gas and its density is not high enough (B13). Accordingly, $C_s$ will likely be pushed to higher values and $f_{des}$ to lower values. Other dust processes such as shattering and PEH can also influence the parameters' range. Shattering will possibly increase both $f_{des}$ and $C_s$ because it enhances dust destruction and growth processes. PEH suppresses SF (Forbes et al. 2016; Hu et al. 2017), and hence this process has the same effect as implementing H$_2$--dependent recipe. We will investigate these different effects in our forthcoming papers.

Table 4 also shows ratios of the molecular hydrogen (M$_{H_2}$/M$_{HI}$, M$_{H_2}$/M$_{\ast}$), dust (M$_{D}$/M$_{Z}$, M$_{D}$/M$_{\ast}$, M$_{D}$/M$_{G}$), and metals (M$_{Z}$/M$_{\ast}$) in all models at T = 1 Gyr. Where $M_{H_2}$, $M_{HI}$, $M_{\ast}$, $M_{D}$, $M_{Z}$, and $M_{G}$ are the total mass of H$_2$, HI, stars, dust, metals, and total gas in each model. Our models sit in two groups according to the observed M$_{H_2}$/M$_{HI}$ ratio in the MW (about 0.25; Yin et al. 2009), models with M$_{H_2}$/M$_{HI}$ $\geq$ 0.25  (namely, M1, M11--M14) and models with M$_{H_2}$/M$_{HI}$ $<$ 0.25 (the rest of the models). This result is a reflection of the efficiency of the transformation of the metal between the dust and gas phases. The same models within the higher range of M$_{H_2}$/M$_{HI}$ ratio are in the higher range of the M$_{D}$/M$_{Z}$ ratio (M$_{D}$/M$_{Z}$ $\geq$ 0.4, where 0.4 is the MW value). The number of models that approximately reproduce the MW  M$_{D}$/M$_{\ast}$ ratio (0.001) is slightly extended to include, along with the models mentioned above, M2, M16, and M21. While about half of the models (M1, M2, M9--M14, M20, and M21) nearly have the MW dust--to--gas ratio (0.01).

\begin{figure*}
\begin{multicols}{2}
    \includegraphics[width=1.\linewidth]{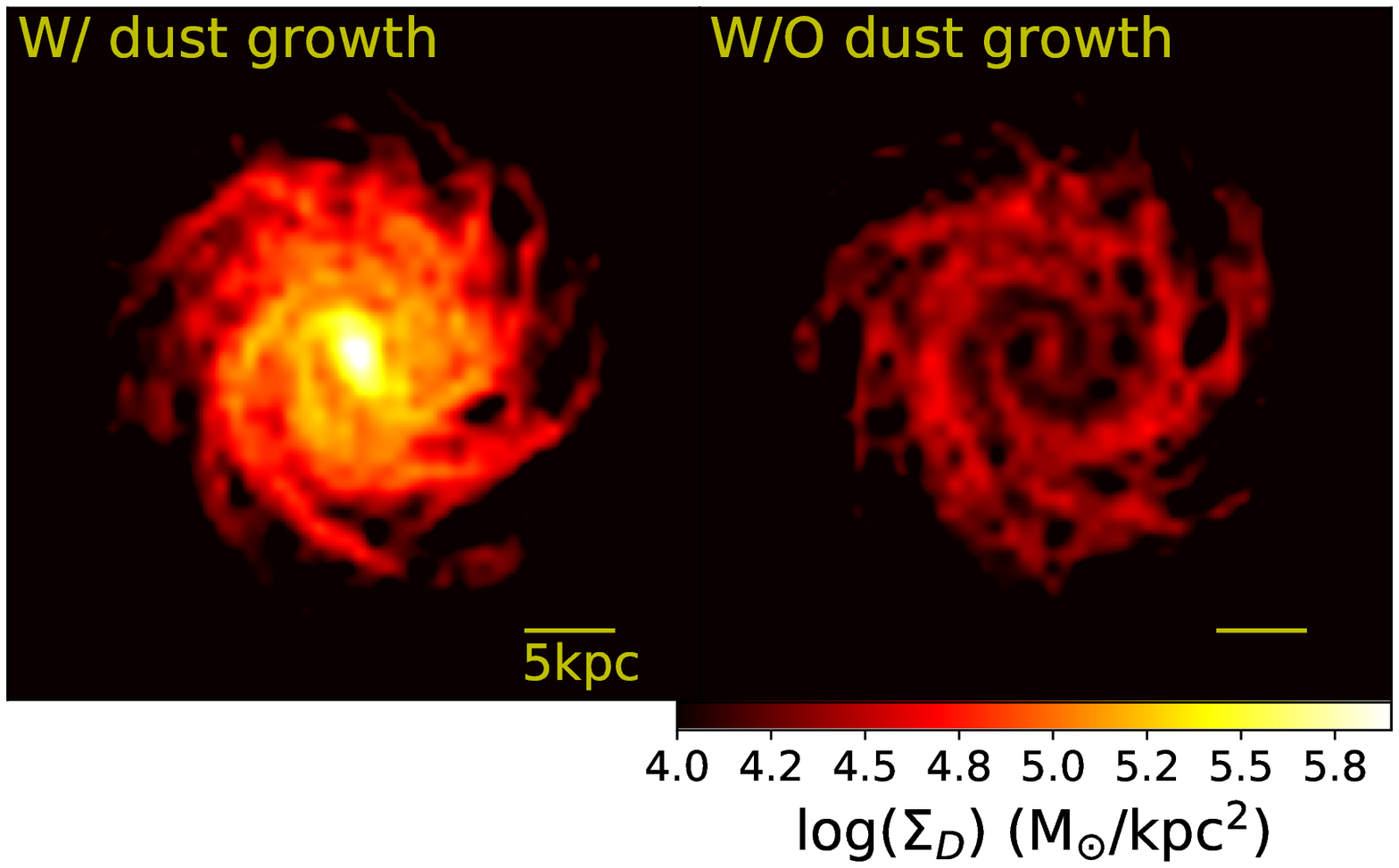}\par 
    \includegraphics[width=1.\linewidth]{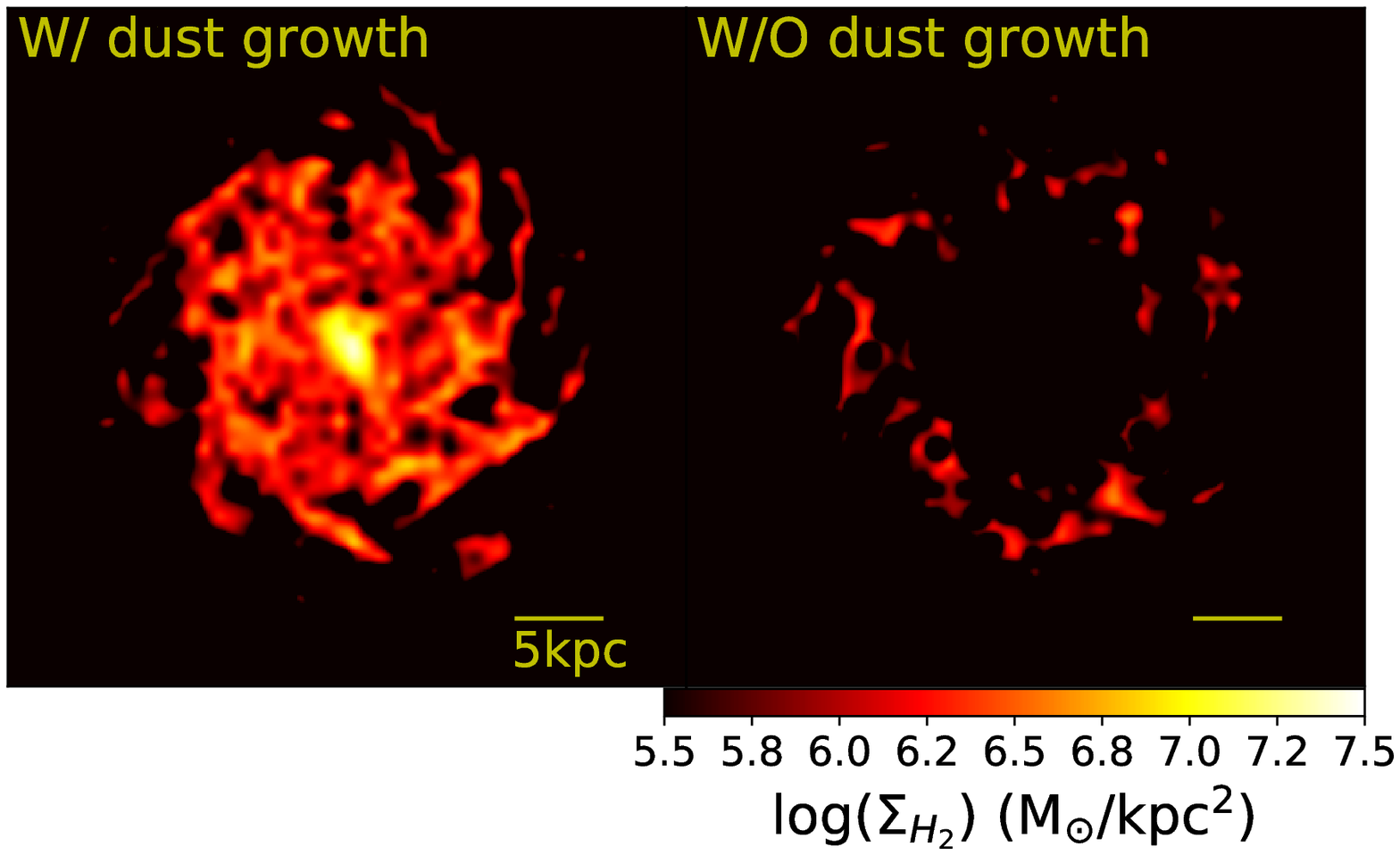}\par 
    \end{multicols}
\begin{multicols}{2}
    \includegraphics[width=1.\linewidth]{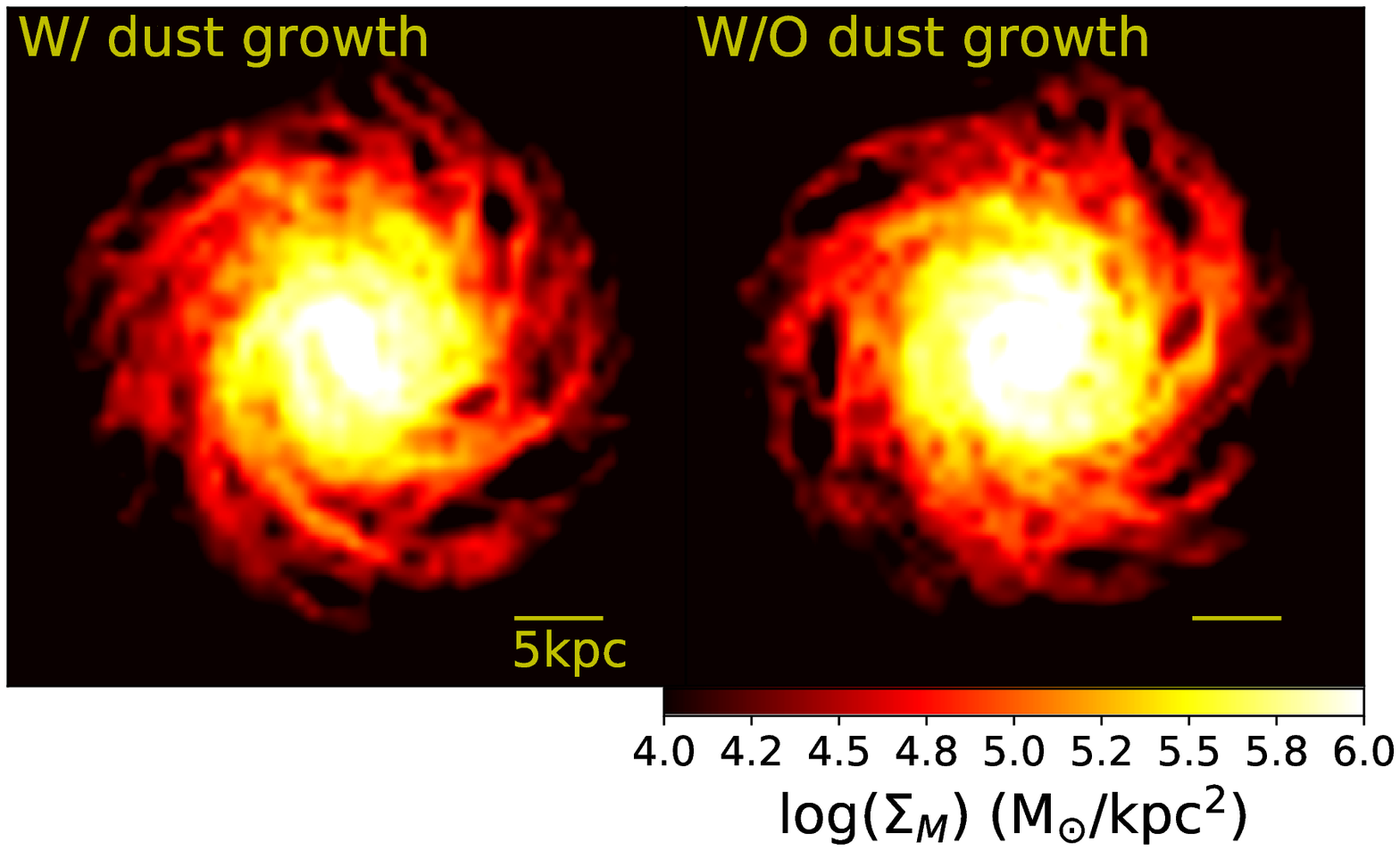}\par 
    \includegraphics[width=1.\linewidth]{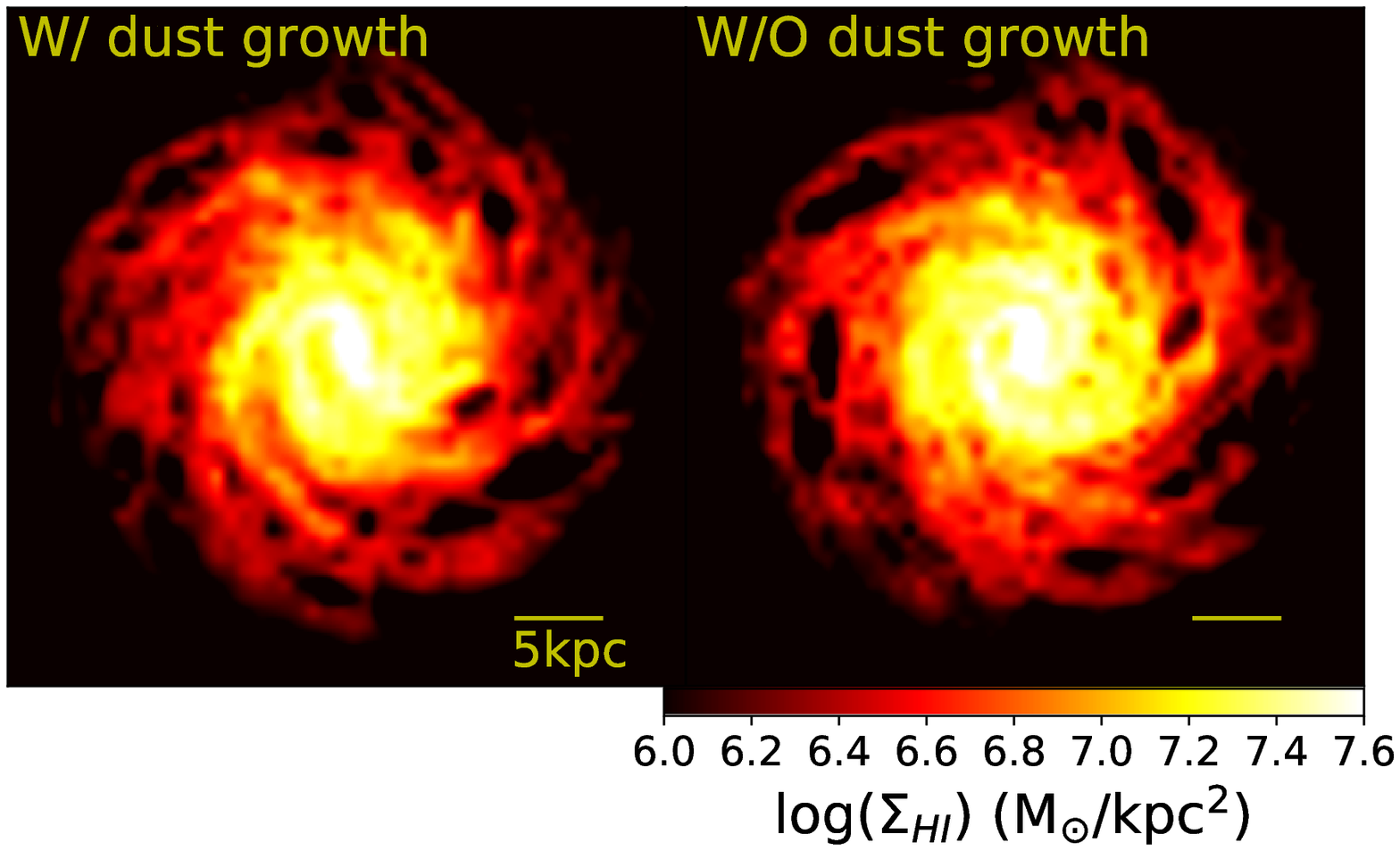}\par 
    \end{multicols}
\caption{Same as in Fig. \ref{fig1} for models M2 (left panel of each subfigure, $f_{des}$ = 0.02, $C_s$ = 1) and M15 (right panel of each subfigure, $f_{des}$ = 0.02, $C_s$ = 0).}
\vspace{-0.3cm}
\label{fig2}
\end{figure*}

\begin{figure*}
\begin{multicols}{2}
     \includegraphics[width=1.\linewidth]{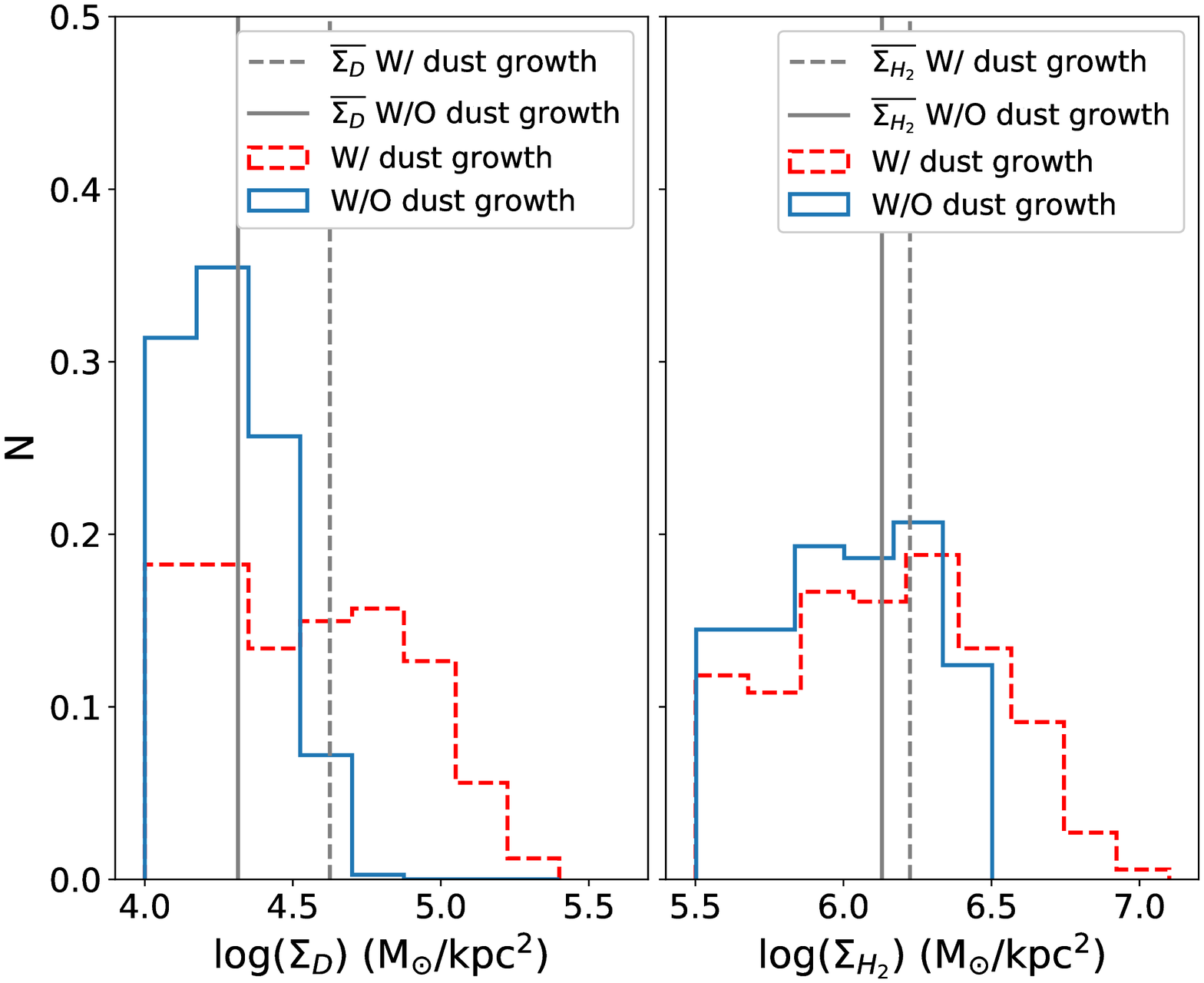}\par 
    \includegraphics[width=1.\linewidth]{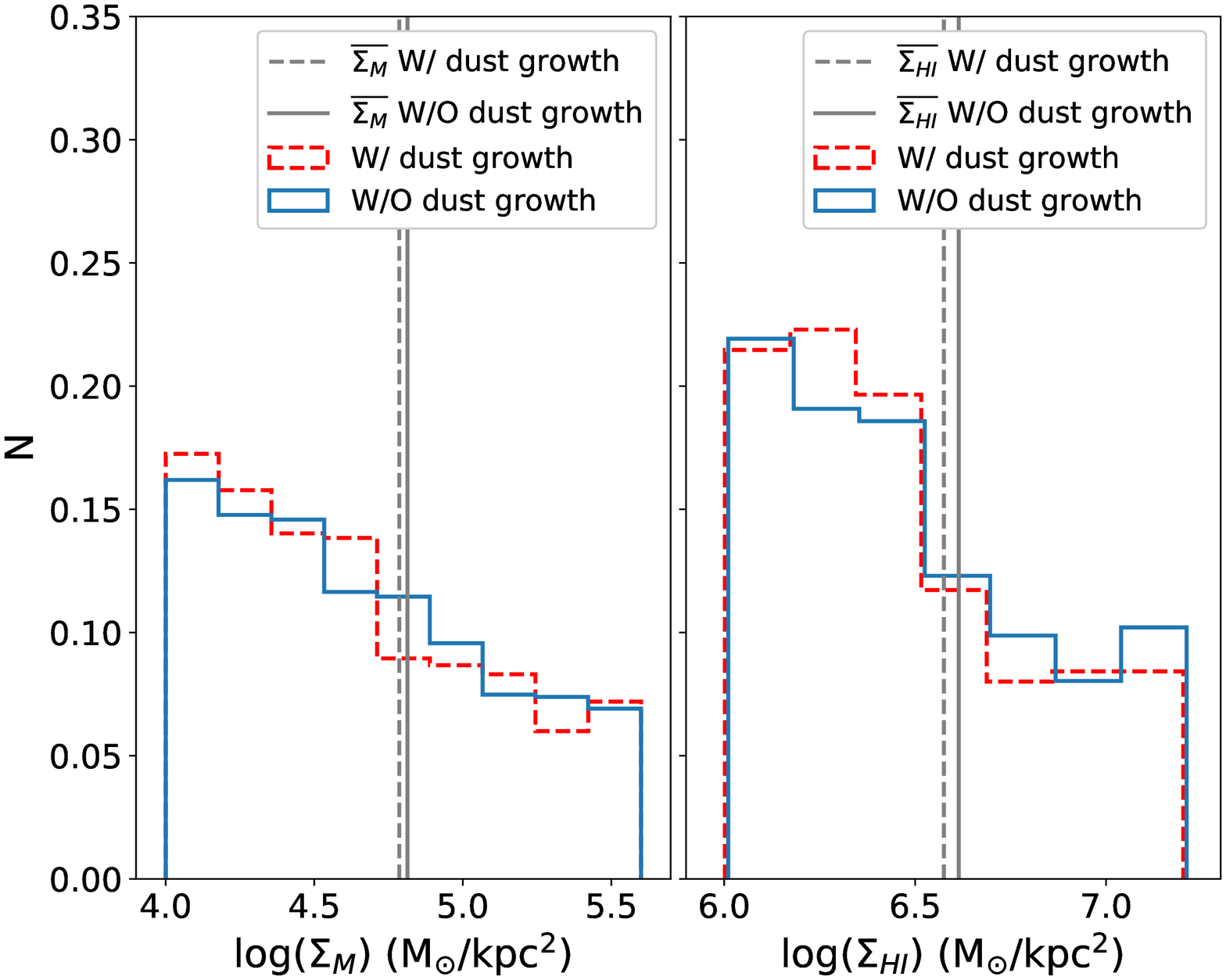}\par 
    \end{multicols}
\caption{Same as in Fig. \ref{hist1} for models M2 (red dashed) and M15 (blue solid).}
\vspace{-0.3cm}
\label{hist2}
\end{figure*} 

\section{Results}
\subsection{Dust and gas spatial distribution}
Fig. \ref{fig1} shows the influence of dust grwoth on the dust (upper left), H$_2$ (upper right), metals (bottom left), and  HI (bottom right) surface density distributions ($\Sigma_D$, $\Sigma_{H_2}$,  $\Sigma_M$, and $\Sigma_{HI}$, respectively) at T = 1 Gyr. Surface densities for the model with dust growth (i.e. $C_s$ has a none zero value; W/) are displayed on the left panel of each subfigure and on the right for the model without dust growth ($C_s$  is zero, i.e. infinite accretion timescale; W/O). Fig. \ref{fig1} shows MW models M1 ($f_{des}$ = 0.01, $C_s$ = 1) and M5 ($f_{des}$ = 0.01, $C_s$ = 0).
\begin{figure}
\begin{center}
\includegraphics[height=6cm,width=8cm]{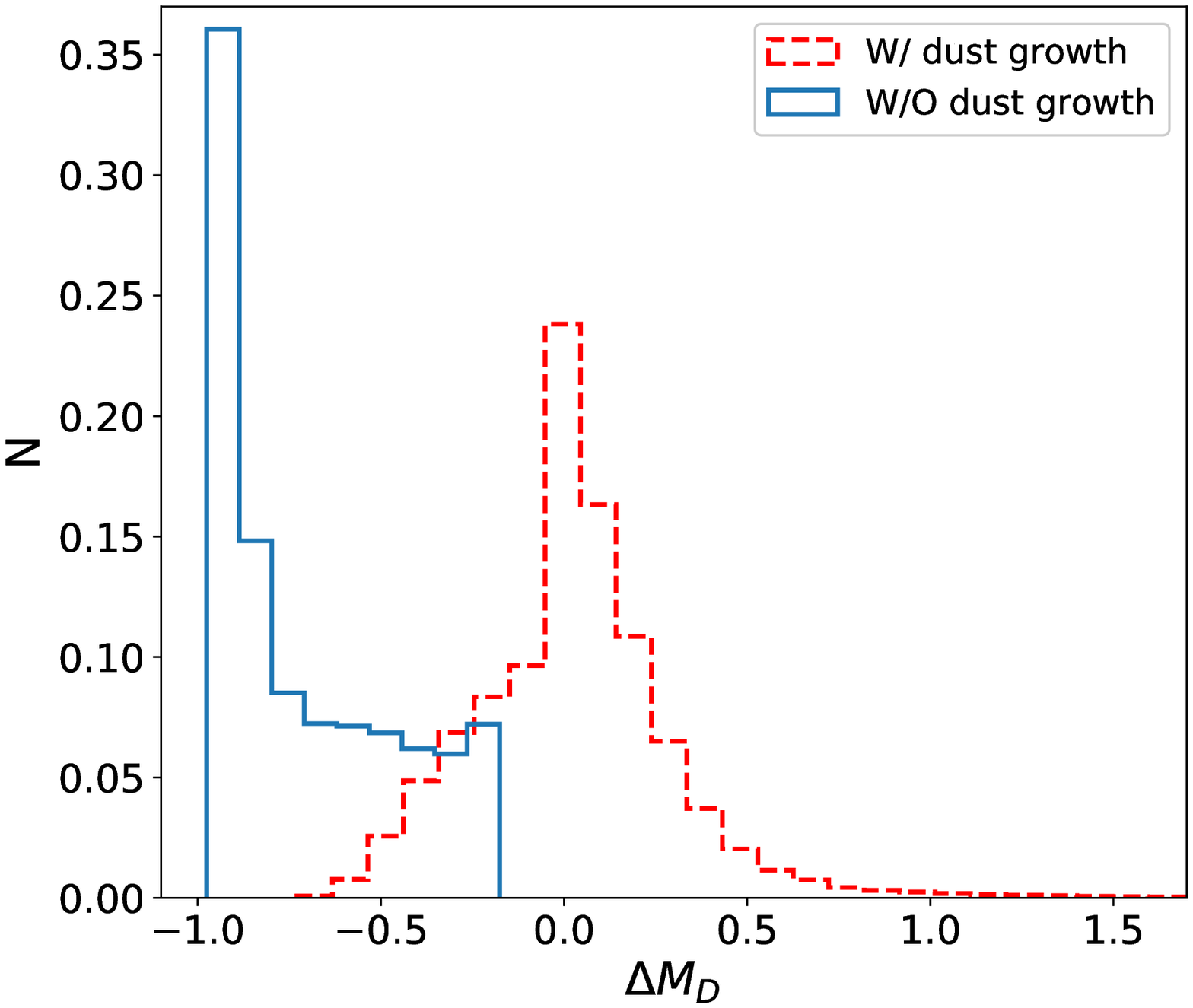}
\end{center}
\caption{Same as in Fig. \ref{hist3} for M2 (red dashed) and M15 (blue solid).}
\label{hist4}
\end{figure} 
\begin{figure*}
\begin{multicols}{2}
    \includegraphics[width=.9\linewidth, height=6cm]{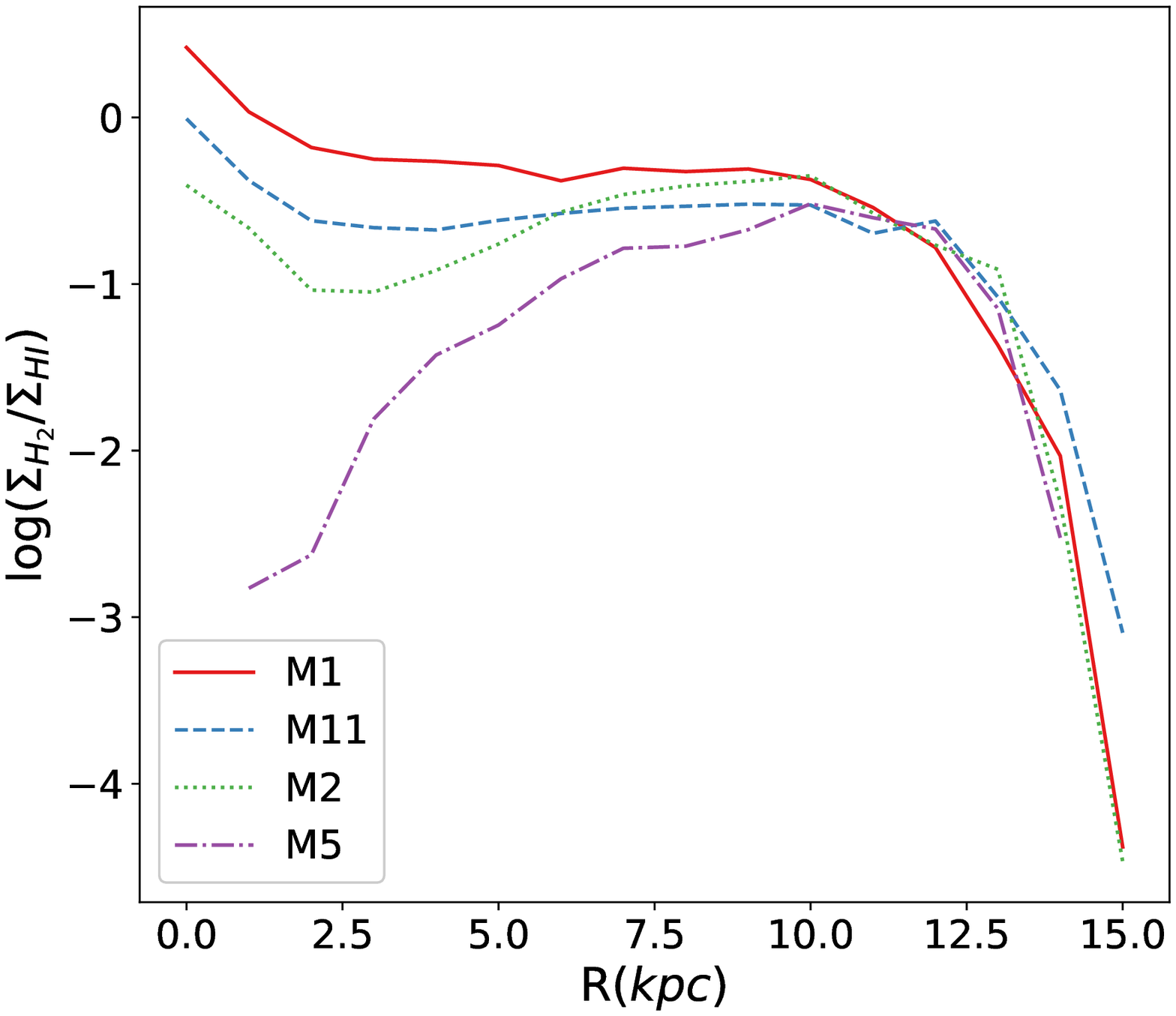}\par 
    \includegraphics[width=.9\linewidth, height=6cm]{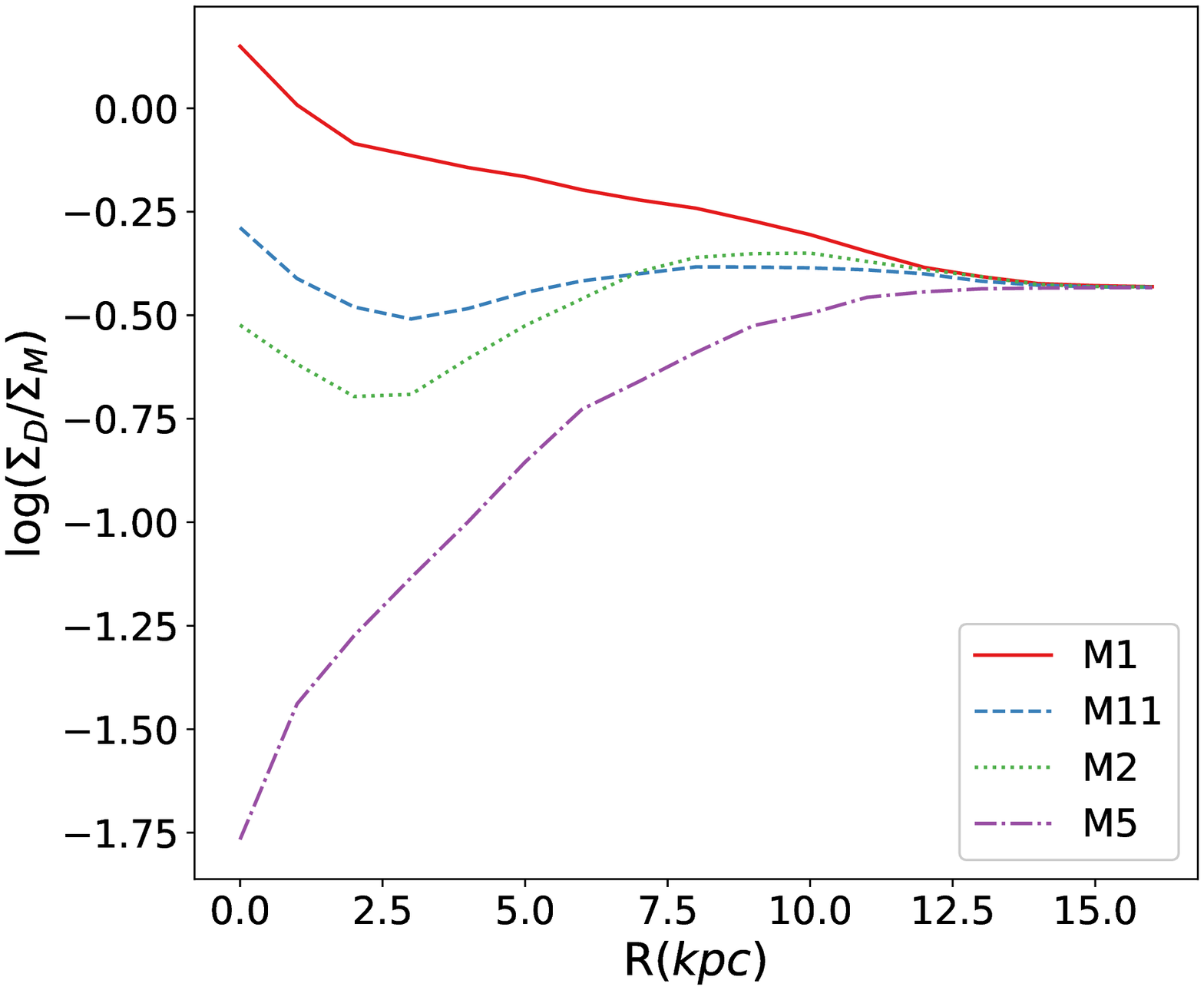}\par 
    \end{multicols}
\caption{Radial profiles of H$_2$--to--HI ($\Sigma_{H_2}$/$\Sigma_{HI}$; left) and dust--to--metals ($\Sigma_{D}$/$\Sigma_{M}$; right) surface density ratios  in lognormal scale, for models M1 (red solid lines), M2 (green dotted lines), M5 (magenta dashed dotted lines), and M11 (blue dashed lines).}
\vspace{-0.3cm}
\label{radial}
\end{figure*}
Dust growth depletes metals from the gas--phase (Savage \& Sembach 1996; Jones 2000; Sofia 2004), accordingly, metals surface density is lower and dust surface density is higher in M1 model compared to M5. Dust growth also enhances HI--to--H$_2$ conversion rate in M1 since molecular hydrogen is efficiently formed on surfaces of dust grains (Fukui \& Kawamura 2010; Cazaux \& Tielens 2004). This leads to a higher abundance of H$_2$ and lower abundance of HI in M1 compared to M5. Moreover, molecular hydrogen has clumpy structure in both models compared to the spiral structure of HI, metals, and dust. This is because H$_2$ is formed in the densest regions of the ISM rather than merely along the spiral arms. 

Furthermore, the presence of a bar in those models enhances star formation in the central regions by channelling the gas inwards (Kormendy \& Kennicutt 2004; Kormendy 2013; Cervantes--Sodi 2017), hence dust formation (destruction) and metal enrichment by stars are enhanced as well.  This enhancement results in the higher densities of dust and metals in the central regions. However, the intense radiation field in those regions combined with the low dust abundance reduce H$_2$ abundance efficiently in the central regions of M5 model. At variance with M1 model where the enhanced H$_2$ formation rate due to the dust growth is able to counterbalance the process. 

Fig. \ref{hist1} shows normalised histograms of the dust (left), H$_2$ (middle left), metals (middle right), and HI (right) surface densities ($\Sigma_D$, $\Sigma_{H_2}$,  $\Sigma_M$, and $\Sigma_{HI}$, respectively) at T = 1 Gyr in models M1(red dashed) and M5 (blue solid). The vertical gray dashed, and solid lines represent the mean of each surface density in M1 and M5, respectively. As discussed above, the model with dust growth (M1) dominates the dust and H$_2$ distributions with higher mean surface densities. The contrary is true for the metals and HI surface densities where the model without dust growth (M5) dominates the distributions with higher mean values, although only slightly higher. For clarity, we also plot the normalised histograms of the dust growth rate in 1 Gyr of evolution ($\Delta M_{D}$) in Fig. \ref{hist3} for the models M1 (red dashed) and M5 (blue solid). $\Delta M_{D}$ is defined as:
\begin{equation}
\Delta M_{D} = \frac{M_D(final) - M_D(initial)}{M_D(initial)}
\end{equation}
where $M_D(initial)$ and  $M_D(final)$ are the initial and final dust mass in each gas particle. In the case of M1 model, almost all the particles have a positive growth rate with a mean close to unity (0.80) contrary to M5 which has a mean of $-0.40$.

Arguments made for the models in Figs \ref{fig1}, \ref{hist1}, and \ref{hist3} are true for the models M2 ($f_{des}$ = 0.02, $C_s$ = 1) and M15 ($f_{des}$ = 0.02, $C_s$ = 0) in Figs  \ref{fig2}, \ref{hist2}, and \ref{hist4}. However, the comparison between each pair of those figures emphasises the significance of the parameter $f_{des}$ in the dust, HI, H$_2$, and metals evolution, $f_{des}$ determines the fraction of dust destroyed in a single SPH particle (higher values of $f_{des}$ corresponds to higher destroyed fractions; see sections 2.2 and 2.7). When $f_{des}$ is doubled from 0.01 in models M1 and M5 to 0.02 in M2 and M15 models, dust total mass dropped by more than a half in model M2 compared to M1 and by about two thirds in model M15 compared to M5. Thus, the molecular hydrogen surface density is higher in models  M1 and M5 compared to M2 and M15 models, while metals and atomic hydrogen densities are higher in M2 and M15. This is because metal incorporation back to the gas--phase from dust phase (dust destruction) is more efficient and HI--to--H$_2$ conversion is less efficient in M2 and M15 models.

Dust is effectively destroyed in spiral arms due to SNe explosions of massive stars therein. Thus, the enhanced dust destruction in Fig. \ref{fig2} diminishes the association of the dust with the spiral arms in M2 model compared to M1. In the extreme case of M15, regions with the lowest dust surface density track the spiral arms precisely. Metals association with the spiral arms is also influenced however less severely. The slightly different structure exhibited by the different models (M1, M2, M5, and M15) indicates that $f_{des}$ and $C_s$ influence the ISM structure as a whole and not only the dust abundance. Furthermore, the higher fraction of dust destroyed causes a significant fraction of the gas particles in M2 model to have negative $\Delta M_{D}$ (Fig. \ref{hist4}). This results in an overlap with M15 model in the range from $-0.12$ to $-0.60$ in which $\Delta M_{D}$ smoothly declines for M2. The mean dust growth rate decreased by order of magnitude in M2 (0.05) compared to M1, and by a factor of 0.14 in M15 ($-0.54$) compared to M5.

\begin{figure*}
\begin{multicols}{2}
    \includegraphics[width=.9\linewidth, height=6cm]{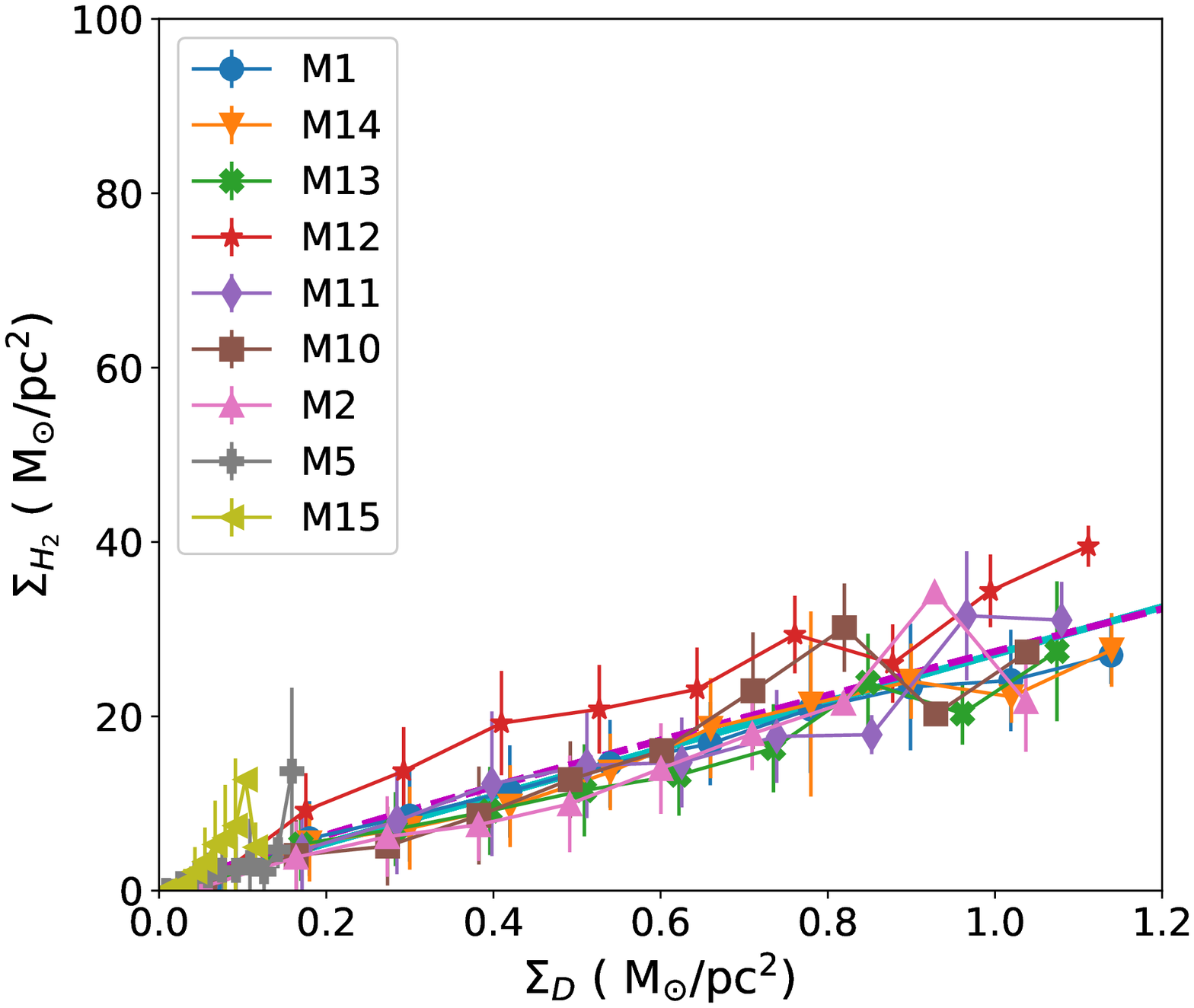}\par 
     \includegraphics[width=.9\linewidth, height=6cm]{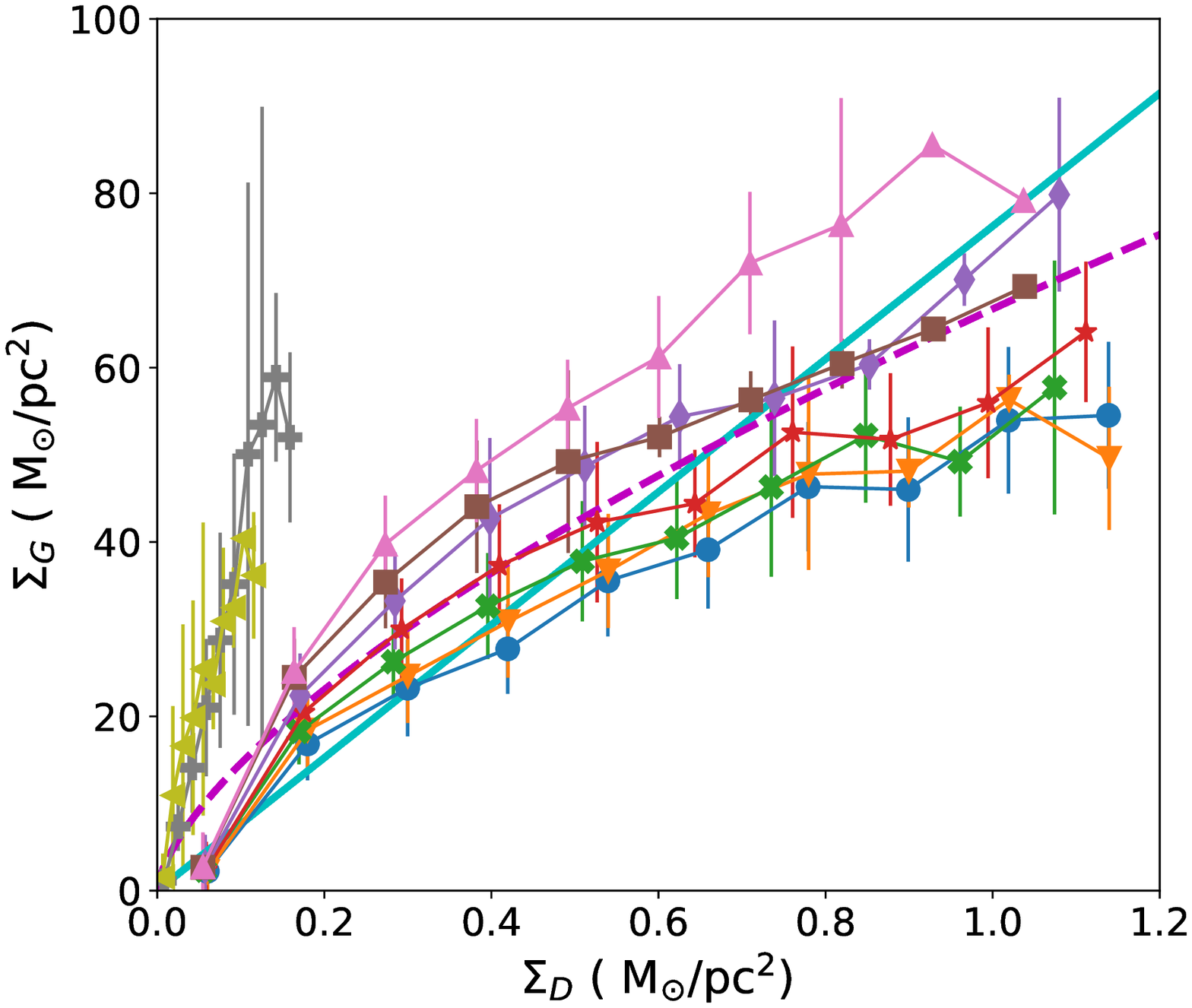}\par 
    \end{multicols}
\caption{H$_2$ surface density versus dust surface density ($\Sigma_{H_2}$ vs  $\Sigma_D$; left) and  total gas surface density versus dust surface density ($\Sigma_{G}$ vs $\Sigma_D$; right) for the models M1, M2, M5, M10--M15 (see Table 4 for the models specifications). Solid lines with different symbols/colours and error bars represent the bin mean and standard deviation of the simulated data in the different models. Cyan solid  and magenta dashed lines are the mean linear and nonlinear fits to the models M1, M2, M10--M14, respectively.}
\vspace{-0.3cm}
\label{fig11}
\end{figure*}
\subsection{Radial profiles}
Fig. \ref{radial} shows the radial profiles of H$_2$--to--HI ($\Sigma_{H_2}$/$\Sigma_{HI}$; left) and dust--to--metals ($\Sigma_{D}$/$\Sigma_{M}$; right) surface density ratios  in lognormal scale, for models M1 (red solid lines), M2 (green dotted lines), M5 (magenta dashed dotted lines), and M11 (blue dashed lines). Models with dust growth (M1, M2, and M11) have similar H$_2$--to--HI profiles, where they steadily decline in the inner 2--3 kpc from 0.4, $-0.02$, and $-0.4$ dex to $-0.3$, $-0.7$, and $-1$ dex, respectively. This decline is followed by almost constant H$_2$--to--HI ratio between 2--3 kpc and 12 kpc in M1 and M11 models before a rapid drop in the outer regions occurs. In M2,  H$_2$--to--HI ratio raises by a factor of 0.2 dex at 6 kpc compared to its value in the inner 3 kpc, and continues to raise to reach a peak at 10 kpc before dropping to $-4.5$ dex in the outer regions. The model without dust growth (M5) starts with much lower H$_2$--to--HI ratio in the inner 2--3 kpc compared to the other models. M5 has H$_2$--to--HI ratio three orders of magnitude lower than M1 in the first kpc. However, it shares a similar trend with M2 in radii $>$ 3 kpc. In all models with $f_{des}$ $<$ 0.02 and $C_s$ $>$ 0.6, H$_2$ dominates the gas surface density in the inner 1 kpc. M11 is the first model in which HI dominates the gas density throughout the galaxy except, perhaps, the innermost region.

Similar to H$_2$--to--HI  profiles, dust--to--metals profiles in models M1, M2, and, M11 gradually decrease from 0.15, $-0.29$, and $-0.50$ dex to $-0.10$, $-0.50$, and $-0.70$ dex, respectively, in the inner 2--3 kpc. Then steadily decline farther in M1, or slightly/significantly raise before flattening in models M11 and M2, respectively. M5 has opposite trend compared to M1, M2, and M11, the dust--to--metals ratio increases continually from $-01.80$ dex at the centre to $-0.40$ dex in the outer skirts of the simulated galaxy. All the models have almost the same dust--to--metals ratio beyond 13 kpc where dust processing is not affecting the initial dust and metals distributions much. M1 and M14 are the only models in which the majority of the metals in the inner 1 kpc are in the dust phase.  In both panels of the H$_2$--to--HI and dust--to--metals profiles, models are ordered in such a way that models with higher dust--to--metals ratio also have higher H$_2$--to--HI ratio.

Dust destruction and growth parameters influence radial profiles as well. Models with low $f_{des}$ (= 0.01) and high $C_s$ ($> 0.6$) produce sufficient dust amount throughout the galaxy. In this case, dust--to--metals ratio maintain exponential profiles and hence H$_2$--to--HI ratio also maintain nearly exponential profiles. On the other hand for $C_s$ $\leq$ 0.6 (and M2), dust--to--metals and H$_2$--to--HI deviate from exponential profiles since the models have a considerable  amount of dust in the outer regions due to the low number of SNe going off therein. This causes dust--to--metals profiles to raise instead of continuing to decline, and then drop beyond 7.5 kpc. The increase of the dust--to--metals ratio combined with the weak stellar radiation field causes H$_2$--to--HI ratio to raise as well before dropping beyond 10 kpc.

\subsection{Dust spatial correlations}
We explore the spatial correlations of dust in a sample of our models considering two types of correlation;
linear correlation:
\begin{equation}
\Sigma_{H_{2}} = \alpha_{H_{2}l} \Sigma_{D},
\end{equation}
\begin{equation}
\Sigma_{G} = \alpha_{Gl} \Sigma_{D},
\end{equation}
and nonlinear correlation:
\begin{equation}
\Sigma_{H_{2}} = \alpha_{H_{2}nl} \Sigma_{D}^{P},
\end{equation}
\begin{equation}
\Sigma_{G} = \alpha_{Gnl} \Sigma_{D}^{P},
\end{equation}
where $P$ is a power index, $\alpha_{H_{2}l}$, $\alpha_{Gl}$, $\alpha_{H_{2}nl}$, and $\alpha_{Gnl}$ are the slopes of the $\chi^2$ fit to the simulated data for the linear and nonlinear fits to $\Sigma_{H_{2}}$ vs $\Sigma_D$ and $\Sigma_{G}$ vs $\Sigma_D$, respectively. All of the fit parameters are also described in Table 3 for clarity. In the former correlation, we assume that $\Sigma_{H_{2}}$--$\Sigma_D$ follow the same relation considered to be between $\Sigma_{G}$ and $\Sigma_D$ (Israel 1997a; 1997b; Leroy et al. 2007; 2009; Gratier et al. 2010; Leroy et al. 2011; Sandstrom et al. 2013), and  in the later we allow the relationship to be different.

Fig. \ref{fig11} shows H$_2$ surface density versus dust surface density ($\Sigma_{H_2}$ vs  $\Sigma_D$; left) and the total gas surface density versus dust surface density ($\Sigma_{G}$ vs $\Sigma_D$; right) for the models M1, M2, M10--M14 (with dust growth) and  M5, M15 (without dust growth; see Table 4 for the models specifications). Solid lines with different symbols/colours and error bars represent the bin mean and standard deviation of the simulated data in the different models. $f_{des}$ in all models, except for M2 and M5 ($f_{des}$ = 0.02), is 0.01. The cyan solid  and magenta dashed  lines are the mean linear ($\alpha_{H_{2}l}$ = 27$\pm$3, $\alpha_{Gl}$ = 76$\pm$29) and nonlinear ($\alpha_{H_{2}nl}$ = 27.4$\pm$4.4, P = 0.90$\pm$0.07; $\alpha_{Gnl}$ = 66.7$\pm$5, P = 0.66$\pm$0.04) fits to the models M1, M2, M10--M14, respectively. Models lie on $\Sigma_{G}$--$\Sigma_D$ plane according to the amount of dust therein,  which depends on the combination of $f_{des}$ and $C_s$  adopted. When $f_{des}$ is high and/or $C_s$ is low, a steep slope in the $\Sigma_{G}$--$\Sigma_D$ relation is obtained, and vice versa. This is because the gas cooling is dependent, in part, on the dust corrected metallicity. M2 (pink up triangle) is the model that produces the least dust, therefore, it has the steepest slope on the $\Sigma_{G}$ vs  $\Sigma_D$ plane. M2 is proceeded by models with $f_{des}$ = 0.01 in increasing order of $C_s$. In the $\Sigma_{H_2}$ vs $\Sigma_D$ plane, the order of models is less obvious.

Furthermore, the correlation between $\Sigma_{H_2}$ and $\Sigma_D$  is tighter than the correlation between $\Sigma_{G}$ and $\Sigma_D$. Besides,  $\Sigma_{H_2}$--$\Sigma_D$ relation can be fitted by linear correlation, while $\Sigma_{G}$--$\Sigma_D$ relation is better fitted by nonlinear correlation. This is important, though not completely surprising since there is an underlying relation between dust and molecular hydrogen and there is no such direct relation between dust and total gas and atomic hydrogen. While Leroy et al. (2011), Hughes et al. (2104), and Relano et al. (2016) found a linear correlation in both relations, B13 results for $\Sigma_{H_2}$ and $\Sigma_D$ relation show a hint to a nonlinear correlation. Trends of the models without dust growth (M5, M15) are also consistent with linear correlation on the $\Sigma_{H_2}$--$\Sigma_D$ plane. This is because the physics of H$_2$ formation and destruction in both types of models (with and without dust growth) is the same (see section 2.3). M5 and M15 models modify the mean values of the fits slightly on $\Sigma_{H_2}$--$\Sigma_D$ plane ($\alpha_{H_{2}l}$ = 27.5; $\alpha_{H_{2}nl}$ = 29.2, P = 0.93). However, they diverge significantly from models with dust growth (M1, M2, M10--M14) on the $\Sigma_{G}$ --$\Sigma_D$ plane pushing up the mean values of the fits ($\alpha_{Gl}$ = 112.9; $\alpha_{Gnl}$ = 139.8, P = 0.73).

\subsection{Dust scaling relations}
\begin{figure}
\begin{center}
\includegraphics[height=6cm,width=8cm]{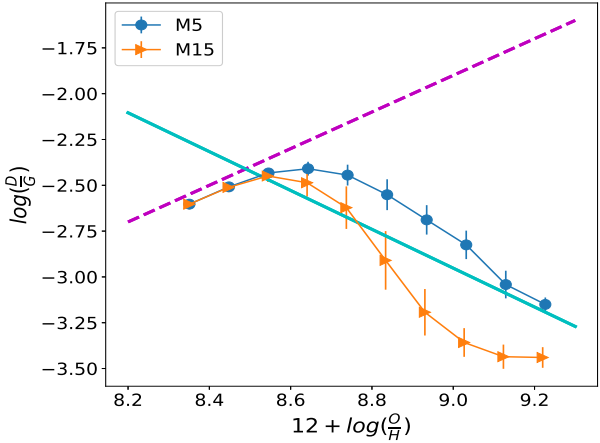}
\end{center}
\caption{$log(D/G)$ versus $12+log(\frac{O}{H})$ (intrinsic metallicity) in models M5 and M15. Solid lines with different symbols/colours and error bars represent the bin mean and standard deviation of the simulated data in the two models. The cyan solid line represents the mean fit to the models, and the magenta dashed line represents the fit to the observations by Remy--Ruyer et al. (2014).}
\label{fig10}
\end{figure}

\begin{figure}
\begin{center}
\includegraphics[height=6cm,width=8cm]{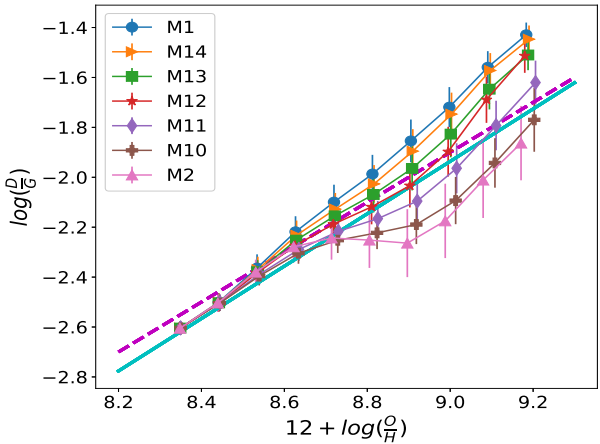}
\end{center}
\caption{Same as in Fig. \ref{fig10} for models M1, M2, M10--M14 where the cyan solid line represents the mean fit to the models ($\beta_{sm}$ = 1.05$\pm$0.3, $b$ = $-11.4\pm$2.4).}
\label{fig6}
\end{figure}

Fig. \ref{fig10} shows  $log(D/G)$ vs $12+log(\frac{O}{H})$ (intrinsic metallicity) in models M5 and M15 (models without dust growth). Solid lines with different symbols/colours and error bars represent the bin mean and standard deviation of the simulated data in the two models. The cyan solid line represents the mean fit to the models, and the magenta dashed line represents the fit to observations by Remy--Ruyer et al. (2014) (the fit parameters to the high metallicity range, $12+log(\frac{O}{H}) > 8.1$, are used). The same is shown in Fig. \ref{fig6} for models M1, M2, M10--M14 where the cyan solid line represents the mean fit to the models with $\beta_{sm}$ = 1.05$\pm$0.3 (the slope of the relation) and $b$ = $-11.4\pm$2.4 (the intersection with $y$--axis). Models with and without dust growth have opposite trends, positive in the former and negative in the later. 

Moreover, below metallicity of 8.42 gas particles have dust--to--metal ratio very close to the initial value in all models, i.e. they are mostly unevolved in terms of dust processing. Once dust processing takes place, the models behave differently according to their adopted $f_{des}$ and $C_s$ values resulting in different patterns. The slope becomes shallower as $f_{des}$ increases and/or $C_s$ decreases accompanied by an increase in the scatter until the trend is reversed when the stellar production fails to counterbalance destruction processes in the lack of or weak dust growth as in models M5 and M15 (see Table 4 for the individual models slopes).


\begin{figure*}
\begin{center}
\includegraphics[width=1.\linewidth]{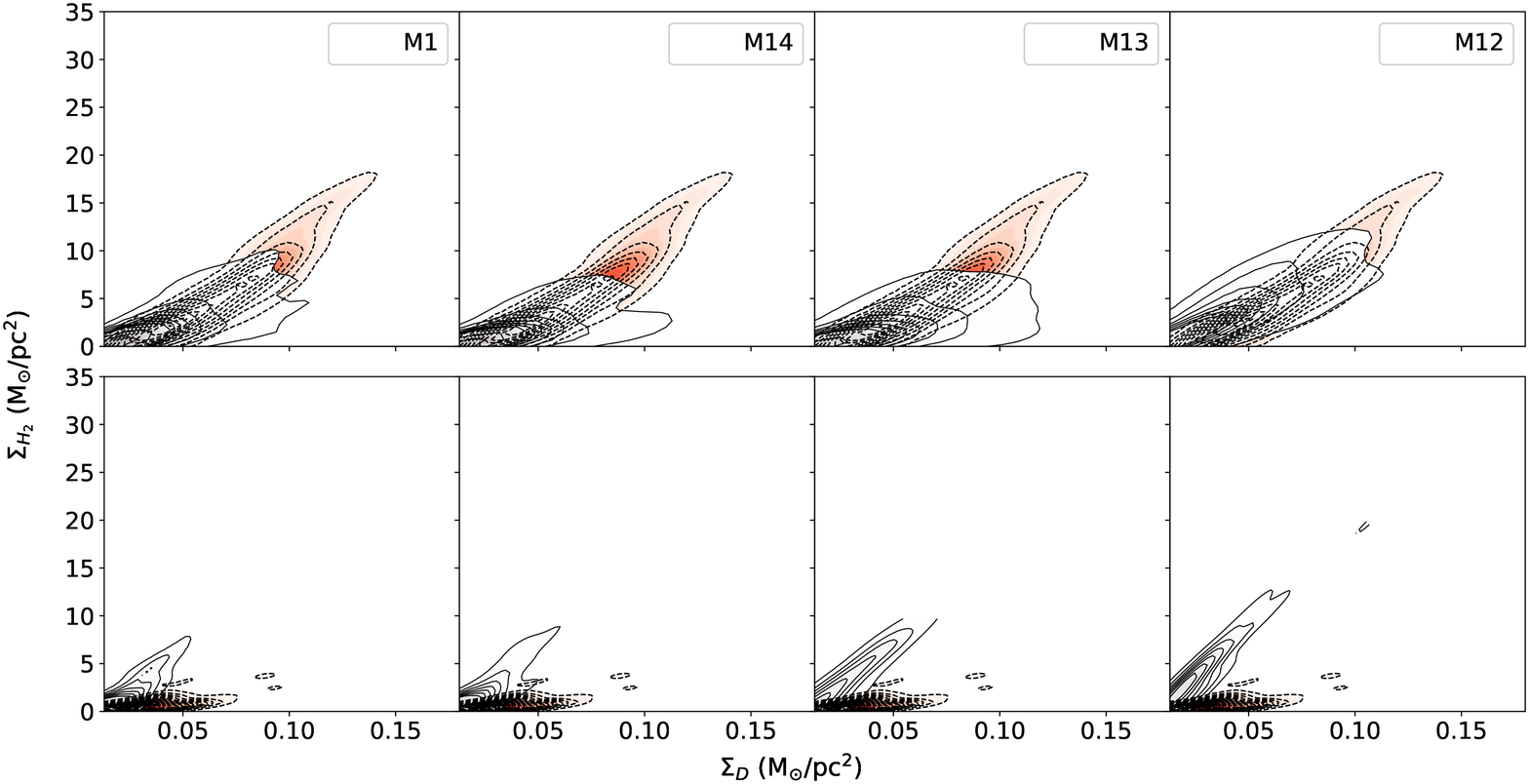}\par
\end{center}
\caption{The 2D distribution of $\Sigma_{H_{2}}$ vs $\Sigma_{D}$ presented by contours. Red filled dashed contours show M101 data, and grey filled solid contours show M1, M14, M13, and M12 models from left to right, respectively. Top and bottom rows include data with $A_O$ $>$ 8.4 and $A_O$ $\leq$ 8.4 (gas--phase metallicity), respectively.}
\label{fig13}
\end{figure*}

Figs \ref{fig10} and \ref{fig6} illustrate the influence of variations of dust parameters ($f_{des}$ and $C_s$) on the time evolution of dust. A unity slope in the $log(D/G)$ vs $12+log(\frac{O}{H})$ relation is obtained only through averaging over a range of dust properties, as shown in Fig. \ref{fig6} by the models’ mean fit (cyan line) and the best fit to observations by Remy--Ruyer et al.  (2014)  (dashed magenta line). Indeed looking at the individual models and their scatter, we find that when $C_s$ is less than or equal 0.7 and/or $f_{des}$ is higher than 0.01, the relation is better fitted by two slopes with a transition around 8.89 dex. In the extreme cases of models M2 and M10, two transitions are present around the solar metallicity and 8.89 dex. This may prompt the need to decently probe both ends of the metallicity range to have a full understanding of this relationship and its consequences concerning galaxy evolution. 

\section{Comparison with M101 galaxy}
Thanks to space facilities such as Herschel and Spitzer the 2D maps of dust and metals now are available with high resolution (e.g. Leroy et al. 2011; Aniano et al. 2012; Sandstrom et al. 2013; Draine et al. 2014; Calzetti et al. 2018; Chiang et al. 2018). In this section, we attempt to compare results of the models with optimal $f_{des}$ and $C_s$ values to the data of the spatially resolved M101 galaxy and discuss the models limitations. This is primarily motivated by the study presented in Chiang et al. (2018) which represents one of the most accurate studies of the resolved dust--to--metal ratio in nearby galaxies with data available to us and not limited to radial profiles (Chiang et al. 2018, private communication). M101 and the MW have similar stellar mass (5.3$\times10^{10}$ M$\sun$ van Dokkunm et al. 2014; 5$\times10^{10}$ M$\sun$ Yin et al. 2009), size ($r_{25}$ = 16.7 kpc Mihos et al. 2013; 16 kpc Bigiel \& Blitz 2012), and gas fraction in the disk (0.3 Walter et al. 2008; 0.2 Yin et al. 2009). The average metallicity of M101, however, is 0.25 dex (8.44 in units of oxygen abundance) lower than the MW one (8.69). 

Although dust properties depend mainly on the underlying ISM properties, environmental processes such as tidal interactions and ram pressure stripping could influence its correlations with the gas (e.g. Bekki 2014; Cortese et al. 2016). The well--known asymmetry of M101 signs to possible tidal interactions with its companions that resulted in tidal stripping of gas from M101 (Mihos et al. 2012; 2013; 2018). M101 is not a standard galaxy in terms of its gas radial profiles where the gas profile (H$_2$ + HI) spans only 0.5 dex from the centre to about 16 kpc. This shallow profile is a result of steeply rising HI profile and steeply decreasing H$_2$ profile. With all of this in mind, we use M101 data not to look for a perfect match, but to investigate potential limitations of our model in reproducing the properties of spiral galaxies.

\subsection{Dust spatial correlations}
We start with inspecting M101 data that is available to us. Dust surface density in M101 has a small range compared to our models ($<$ 0.2 M$\sun$/pc$^2$), and it has a bimodal distribution. A large fraction (70\%) of the ISM gas has $\Sigma_{D}$ $<$ 0.07 M$\sun$/pc$^2$ with a peak around 0.03 M$\sun$/pc$^2$ (most of it resides at distances larger than 6 kpc from the centre). Smaller fraction of gas (30\%) has $\Sigma_{D}$ $>$ 0.07 M$\sun$/pc$^2$ with a peak around 0.09 M$\sun$/pc$^2$. Only 23\% of M101 data is for regions within the inner 6 kpc. The wide metallicity range in M101 (7.8 to 8.7 in $12+log(\frac{O}{H})$ units) is mainly caused by the ISM beyond 6 kpc in which metallicity spans 0.71 dex. Additionally, the dust--to--gas ratio in logarithmic scale covers a wide range from $-4.4$ to $-1.9$. Furthermore, looking at the full data of M101 in comparison to the full data of the models, we find that, unlike our models, $\Sigma_{H_2}$--$\Sigma_{D}$ relation in M101 is not accurately described by linear correlation ($\alpha_{H_{2}l}$ = 83.8$\pm$0.84; $\alpha_{H_{2}nl}$ = 1216$\pm$49.5, P = 2$\pm$0.02). However, the bulk of $\Sigma_{G}$--$\Sigma_D$ relation is well described by linear correlation as our models in this range of $\Sigma_{D}$ ($\alpha_{Gl}$ = 208.7$\pm$1.14; $\alpha_{Gnl}$ = 40$\pm$1.6, P = 0.4$\pm$0.013; these values are for the full data sample). 
\begin{figure*}
\begin{center}
\includegraphics[width=1.\linewidth]{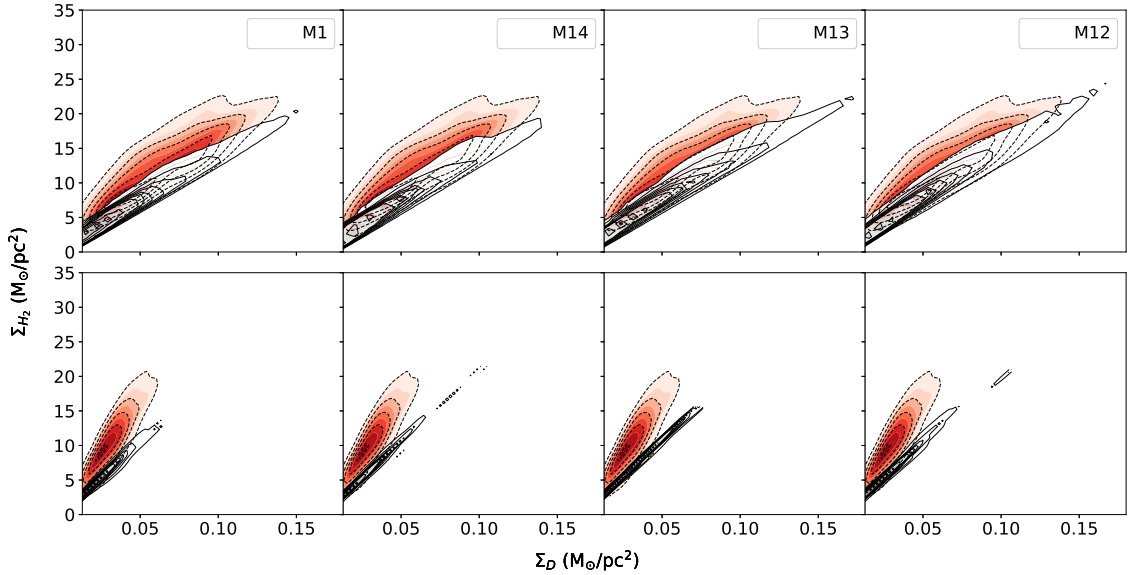}
\end{center}
\caption{Same as in Fig. \ref{fig13} for $\Sigma_{G}$ vs $\Sigma_{D}$.}
\label{fig14}
\end{figure*} 

In the following, we try to compare regions in M101 and the models that have similar ISM properties instead of comparing the full data. For that purpose, we use the dust--to--gas ratio and metallicity as indicators of the ISM properties. First, we select all the ISM regions (in M101 and the models) with dust--to--gas ratio between the minimum accessible by the models and the maximum in M101 galaxy. This selection excludes regions with low dust--to--gas ratio in M101 (below the minimum value accessible by the models) and regions with high dust--to--gas ratio in the models (beyond M101 maximum value). Furthermore, we discard all the regions with dust--to--metals ratio close to the initial value (within an error of 0.05) and located beyond 13 kpc from the centre of the models since they are mostly unevolved in terms of dust processing. These regions have dust surface density below the minimum in M101 data from the previous step. The data selected in the previous steps is then split into two bins according to the gas--phase metallicity, 8.2 $\leq$ $A_O$ $<$ 8.4 and 8.4 $\leq$ $A_O$ $\leq$ 8.7 ($A_O$ $\equiv$ $12+log(\frac{O}{H})$), hereafter low and high metallicity bins, respectively. Switching to the gas--phase metallicity shifts the metallicity range in the models by about 0.14 dex to the left, i.e. regions that we consider unevolved in terms of dust processing now have metallicity below 8.26. Note that the data used in and excluded from the comparison in both M101 and models overlap in the gas surface density versus metallicity plane by various degrees. This selection, particularly, addresses the mutual dependence between dust and H$_2$ ($\Sigma_{H_2}$--$\Sigma_{D}$) since the molecular hydrogen fraction is a function of the dust--to--gas ratio along with other ISM properties in the present simulation (see section 2.3).

Fig. \ref{fig13} shows the 2D distribution of $\Sigma_{H_{2}}$ vs $\Sigma_{D}$ presented by contours. Red filled dashed contours show M101 data, and grey filled solid contours show M1, M14, M13, and M12 models.  The top row shows the results for the high metallicity bin while the bottom row shows results for the low metallicity bin. In the top row, $\Sigma_{H_{2}}$--$\Sigma_{D}$ correlation in M101 is reproduced well by the models up to dust surface density slightly above 0.1 M$\sun$/pc$^2$ despite the higher scatter seen in the models. At this dust surface density, the density of the data is two, four, and more than four times lower in models M12, M1, M13, and M14, respectively, compared to the density in M101.  Around the origin, the models have two times higher density on average. As $\Sigma_{D}$ increases ($>$ 0.1 M$\sun$/pc$^2$) the location of the ISM gas in M101 and in the models shifts towards the central regions. In these regions, M101 maintains high H$_2$ surface density that increases quadratically with  $\Sigma_{D}$,  while the intense radiation field in the models suppresses H$_2$ surface density compared to M101. Only when $\Sigma_{D}$ is higher than 0.4 M$\sun$/pc$^2$, $\Sigma_{H_{2}}$ can reach values as high as in M101. To understand the difference between our models and M101 data in terms of radiation fields induced by stars, we compare the SFR surface density radial profiles in the models to M101 radial profile (Chiang et al. 2018; Leroy et al. 2019  private communication). The SFR radial profile of M101 is considerably shallower than the models' profiles, however, the central SFR density of the models is higher by about 0.4 dex. Accordingly, H$_2$ in the central regions is subjected to more intense radiation fields in the models.

In the low metallicity bin (bottom row), the models overestimate M101 data. The models produce a significant amount of H$_2$ in these outer regions (11--17.3 kpc) as addressed in section 3.2 due to the reduced radiation field combined with the amount of dust therein. In this low metallicity range, M101 contours reveal two populations. The first with high H$_2$ abundance includes ISM that follows the same trend as in the high metallicity bin (top row), while the second with lower H$_2$ abundance includes ISM with extremely high total gas surface density for its location and metallicity. The gas in these ISM clouds is dominated by HI and it is highly enriched with dust, this causes the deviation from the trend in the high metallicity bin. The outermost contour represents density three times higher than in the top row. Fig. \ref{fig14} is same as Fig. \ref{fig13} for $\Sigma_{G}$ vs $\Sigma_{D}$. M101 has a wide parameter space on this plane with a wing--like structure formed by the low metallicity ISM. The models can hardly reproduce the full parameter space and only trace the bulk of M101 data in the high metallicity bin which contains the bulk of all the data (top row). M101 data in the bottom row is underestimated. The two metallicity bins overlap regarding the  dust--to--gas ratio range. While the overlap is minimal in the models, the high metallicity bin in M101 fully overlaps with the low metallicity bin. This drives the, apparently, worse reproduction of M101 data on this plane.

The selection criterion we used, based on the dust--to--gas ratio and metallicity, could exclude interesting parts of the ISM such as regions with similar gas surface density and metallicity but with higher or lower dust surface density. Hence, a selection based on the gas density and metallicity is ideal, however, it is more challenging because of the unconstraint absolute scale of metallicity. Thus, we made a selection on the gas density to look for those regions. We selected all regions in the ISM with a gas surface density below 20/25 M$\sun$/pc$^2$ for M101 (89\%/97\% of the data) and the models (93\%/95\% of the data). Then this data is split into the same metallicity bins as previously. The results of this selection are not considerably different from the results we described above in the high metallicity bin. In the low metallicity bin, the models only slightly overestimate H$_2$ abundance. This improvement is opposed by the worse results for the total gas abundance. In case the unevolved particles were included in the first selection, the results of the second selection are largely recovered in the low metallicity bin for the models. Accordingly, the two selection criteria yield the same conclusions qualitatively. 

\begin{figure*}
\begin{center}
\includegraphics[height=8cm,width=18cm]{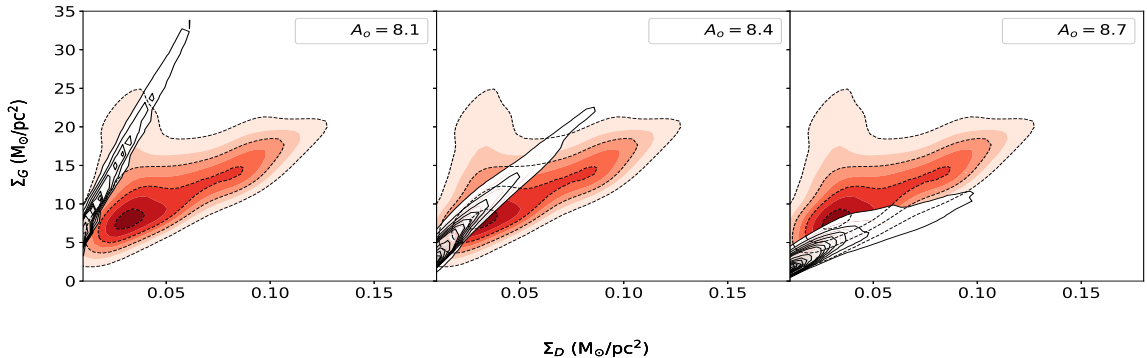}
\end{center}
\caption{The 2D distribution of $\Sigma_{G}$ vs $\Sigma_{D}$ presented by contours. Red filled dashed contours show M101 data, and grey filled solid contours show models with central metallicity $A_O$ = 8.1, 8.4, and 8.7 (MW value) from left to right, respectively.}
\label{fig16}
\end{figure*}

In our aim to understand the wide parameter space in the $\Sigma_{G}$ vs $\Sigma_{D}$ plane especially  the `wing' structure,  we run additional models with lower central metallicities shown in  Fig. \ref{fig16}. Fig. \ref{fig16} shows the 2D distribution of $\Sigma_{G}$ vs $\Sigma_{D}$ presented by contours. Red filled dashed contours show M101 data, and grey filled solid contours show models with central metallicity $A_O$ = 8.1, 8.4, and 8.7 (MW value) from left to right, respectively. There is no selection criterion applied to the data shown. We adopt M1 dust parameters in these models ($f_{des}$ = 0.01 and $C_s$ = 1). While the model with $A_O$ = 8.1 reproduces the upper wing (M101 data with metallicity $\leq$ 8.4), the models with $A_O$ = 8.4 and 8.7 encapsulate the lower wing (M101 data with metallicity $>$ 8.4). We run other models where we changed other parameters such as the metallicity gradiant and star formation threshold density. The results of all the models we ran point towards the difficulty of reproducing the full parameter space of M101 galaxy with our isolated disk models.

\section{Discussion and conclusions}
In this work, we have used our original chemodynamical code to study the influence of dust parameters ($f_{des}$ and $C_s$)  variation on dust and gas spatial correlations, and dust--to--gas ratio versus metallicity relation. $f_{des}$ is a parameter that determines the fraction of dust destroyed in a single SPH particle surrounding SN, higher values of $f_{des}$ corresponds to  higher destroyed fractions. The total mass of dust destroyed by SN is estimated using Nozawa et al. (2006) formulae for dust destruction efficiency and swept mass of the ISM by SN (Eq. A3 and Table 8, and Eq. A4, respectively, see section 2.2 for more details).  $C_s$ is the probability that a metal atom or ion sticks to the dust grain after colliding, i.e., the sticking coefficient. We have also attempted to constrain the possible range of those parameters and compare our models to the spatially resolved M101 galaxy. In the ISM of galaxies, variations in $f_{des}$ and $C_s$ are driven by variations in dust composition (is not resolved here) and ISM conditions. In the present work, the influence of $f_{des}$ and $C_s$ variation is studied using models with different combinations of $f_{des}$ and $C_s$, our findings are summarised in the following.

(1)  The dust--to--gas ratio versus metallicity relation is found to be affected by the variations in $f_{des}$ and $C_s$ since both influence dust abundance in the different epochs of the galaxy evolution. At each metallicity value, more dust is produced when $f_{des}$ is lower and/or $C_s$ is higher in the different models. This leads to a steeper correlation between the dust--to--gas ratio and metallicity with less scatter in each metallicity bin. We also acknowledge the different patterns of the relation in the different models, which are due to variations in $f_{des}$ and $C_s$. Moreover, the linear trend between the dust--to--gas ratio and metallicity consistent with observations  (Remy--Ruyer et al. 2014) is obtained through averaging over models with different dust parameters within the predicted range. The predicted range of $f_{des}$ and $C_s$ is obtained after discarding models that do not yield physical trends of the dust--to--gas ratio versus metallicity relation (negative slopes) as well as models that have slopes below 0.55 (Sandstrom et al. 2013). Repeating this practice for all the models, we identify a final models' set that contains 7 models out of 21 starting models. In this sample, $f_{des}$ ranges between 0.01 and 0.02, and $C_s$ between 0.5 and 1.

According to  the $f_{des}$ values identified, the total fraction of dust destroyed by SN (the mass ratio between the dust destroyed to the dust originally present in all neighbouring SPH particles) ranges on average between 0.42 and 0.44 and the average pre--shock density ranges between 5 and 8 $cm^{-3}$, these results also averages over dust composition (see Table 2 in Nozawa et al. 2006). In low--density media, namely $n_H = 0.25$ $cm^{-3}$, about 25\% of dust is destroyed in our models which is cosistent with the range found in previous studies for silicate and carbonaceous dust (10\%--38\% and 7\%--91\%, respectively, see Table 2 in Micelotta et al. 2018). It is also worth noting that $f_{des}$ value controls whether or not the total mass of dust destroyed by a SN is the same as Nozawa's. Only when $f_{des}$ is 0.02 the total amount of dust destroyed equals to Nozawa's otherwise it is either smaller ($f_{des} < 0.02$) or higher ($f_{des} > 0.02$). Hence, our models favour less dust destruction compared to Nozawa's since most of our feasible models have $f_{des}$ = 0.01 (only one model has $f_{des}$ = 0.02 and none have $f_{des}$ above 0.02). The typical range of  $C_s$ obtained is also within the range commonly adopted/investigated in literature (Spitzer 1978; Leitch-Devlin \& Williams 1985; Hirashita \& Kuo 2011; Zhukovska et al. 2016 and references therein).

(2) $f_{des}$ and $C_s$ variations influence not only the dust abundance but also the abundance of the different ISM components, and the resulting correlations between the cold gas (total and H$_2$) and dust surface densities depend on the $f_{des}$ and $C_s$ adopted values. The positive correlation between $\Sigma_{G}$ and $\Sigma_{D}$ becomes shallower as $f_{des}$ decreases and/or $C_s$ increases but the relation between dust parameters and $\Sigma_{H_2}$--$\Sigma_{D}$ correlation is less obvious. We also find that, while $\Sigma_{H_{2}}$--$\Sigma_{D}$ is well described by linear correlation (as in, e.g., Leroy et al. 2011; Hughes et al. 2104; Relano et al. 2016), $\Sigma_{G}$--$\Sigma_{D}$ (contrary to observations and B13) is better explained by nonlinear correlation. Dust correlation with H$_2$ is found to be tighter than with the total gas reflecting the mutual dependence between dust and H$_2$. However, dust correlations discussed in 3.3 and 3.4 sections possibly evolve with time; accordingly, we run M1 model for 3 Gyr. After 3 Gyr, the slope and intersection of the dust--to--gas ratio versus metallicity relation slightly changed (10\% and 7\%, respectively) while the linear relations between $\Sigma_{H_{2}}$ and $\Sigma_{D}$, and between $\Sigma_{G}$ and $\Sigma_{D}$ became 41\% steeper and 23\% shallower, respectively. The relation between $\Sigma_{H_{2}}$ and $\Sigma_{D}$ is still well described by the linear fit while $\Sigma_{G}$ and $\Sigma_{D}$ relation after 3 Gyr is better described by linear correlation. Radial profiles of the H$_2$--to--HI and dust--to--metals ratios also change according to the adopted $f_{des}$ and $C_s$ values.

(3) Although ISM conditions are the main drivers of dust properties, environmental processes influence its correlations with the gas (Bekki 2014; Cortese et al. 2016). M101 galaxy could be an example where tidal interactions are affecting the gas content (Mihos et al. 2012; 2013; 2018). Accordingly, comparing M101 data to the models without applying any selection criterion reveals the difficulty of reproducing the $\Sigma_{H_{2}}$--$\Sigma_{D}$ correlation despite the bulk of the correlation between $\Sigma_{G}$ and $\Sigma_{D}$ is reproduced. Therefore, we apply a selection criterion that particularly addresses the H$_2$ formation model, and hence the correlation between $\Sigma_{H_{2}}$ and $\Sigma_{D}$ (see section 4). M12 model ($f_{des}$ = 0.01, $C_s$ = 0.7) is the model that best reproduces M101 data according to the selection criterion, however, it underestimates the correlation between dust and H$_2$ in regions with $\Sigma_{D}  >$ 0.1 where the intense radiation field dissociates the modelled H$_2$ efficiently. In the outer regions (metallicity $<$ 8.4), M12 overestimates the correlation. Our simulations are also not able to reproduce the full parameter space on the $\Sigma_{G}$--$\Sigma_{D}$ plane. Models with higher gas fraction do not change these arguments, however, they improve the producibility of the data. After testing several models with different star formation recipes, gas fractions, central metallicities, and metallicity gradients against M101 data, we conclude that (i) metallicity is the primary driver of the spatial dust variations and (ii) Dust--to--gas ratio is the driver of the cold gas spatial variations since it controls the HI--to--H$_2$ conversion rate.

The results we are presenting here are results of simulations with a resolution of $m_{\rm g}=3\times10^4 {\rm M}_{\odot}$ which is indeed not enough to resolve the high--density regions consistent with the characteristic growth timescale reported in  Hirashita (2000) ($10^2$--$10^3$ cm$^{-3}$). Only about 1\% of the gas particles have a density in the density range between $10^2$ and $10^3$ cm$^{-3}$, while the vast majority have densities between 1 and $10^2$ cm$^{-3}$. This would have effects on $f_{des}$ and $C_s$ values that reproduce the observational trends. Accordingly, we also run M1 and M2 models for 1 Gyr with a resolution of 6$\times$10$^5$ M$_{\sun}$.  The amount of dust produced in 1 Gyr by M1 model ($f_{des}$ = 0.01, $C_s$ = 1) was barely changed (identical to the third decimal) and the slope of dust--to--gas ratio vs metallicity relation increased by about 12\%. However, dust produced by M2 ($f_{des}$ = 0.02, $C_s$ = 1) increased by 35\%, which considerably influenced the slope of dust--to--gas ratio vs metallicity relation (increased by about 56\%). Spatial correlations are also influenced in such a way that the relation between $\Sigma_{H_{2}}$ and $\Sigma_{D}$ is 57\% steeper in M1 and M2 low--resolution models than in the original M1 and M2 models. M1 and M2 low--resolution models fall between M1 and M2 models in the $\Sigma_{G}$--$\Sigma_{D}$ plane and their relation can be better fitted with linear correlation contrary to M1 and M2.

\section*{Acknowledgement}
We are grateful to the referee for the constructive and useful comments that improved this paper. We are also grateful to I--DA Chiang for making M101 galaxy data available to us. OO is a recipient of an Australian Government Research Training Program (RTP) Scholarship. LC is the recipient of an Australian Research Council Future Fellowship (FT180100066) funded by the Australian Government.

\section*{References}

\noindent
Andersen M. et al., 2011, ApJ, 742, 7

\noindent
Aniano G. et al., 2012, ApJ, 756, 138

\noindent
Aoyama S. et al., 2017, MNRAS, 466, 105

\noindent
Aoyama  S. et al., 2018, MNRAS, 478, 4905

\noindent
Asano R. S. et al., 2013, MNRAS, 432, 637

\noindent
Asano R. S. et al., 2013, Earth, Planets and Space, 65, 213

\noindent
Barlow, M. J., 1978, MNRAS, 183, 367

\noindent
Bekki K., 2013, MNRAS, 432, 2298; B13

\noindent
Bekki K., Shigeyama T., Tsujimoto T., 2013, MNRAS, 428, L31

\noindent
Bekki K., 2014, MNRAS, 438, 444

\noindent
Bekki K., 2015, 2015, MNRAS, 449, 1625

\noindent
Bertoldi F. et al., 2003, A\&A, 406, L55

\noindent
Bianchi S., Schneider R., 2007, MNRAS, 378, 973

\noindent
Bigiel F., Blitz L., 2012, ApJ, 756, 18

\noindent
Bocchio M, Jones A. P., Slavin J. D., 2014, A\&A, 570, A32

\noindent
Bruzual G., Charlot S., 2003, MNRAS, 344, 1000

\noindent
Calzetti D. et al., 2018, ApJ, 852, 25

\noindent
Cazaux S., Tielens A. G. G. M., 2004, ApJ, 604, 222

\noindent
Cervantes--Sodi B., 2017, ApJ, 835, 80

\noindent
Chiang I.-Da et al., 2018, ApJ, 865, 23

\noindent
Cortese L. et al., 2016, MNRAS, 459, 3574

\noindent
Croxall K. V. et al., 2016, ApJ, 830, 4

\noindent
Dell’Agli F. et al., 2017, MNRAS,  467, 4431

\noindent
Dell’Agli F., 2017, Memorie della Societa Astronomica Italiana, v.88, p.383

\noindent
De Looze I. et al., 2017, MNRAS, 465, 3309

\noindent
Draine B. T., 2003, ARA\&A, 41, 241

\noindent
Draine B. T., 2007, ApJ, 663, 894

\noindent
Draine B. T. et al., 2014, ApJ, 780, 172

\noindent
Dulieu F. et al., 2013, Sci. Rep., 3, 133

\noindent
Dwek E., Scalo J. M., 1980, ApJ, 239, 193

\noindent
Dwek E., 1998, ApJ, 501, 643; D98

\noindent
Dwek E., Galliano F., Jones A. P., 2007, ApJ, 662, 927

\noindent
Ferrarotti A. D., Gail H. P., 2006,  A\&A, 553, 576

\noindent
Forbes J. C. et al., 2016, Nature, 535, 523

\noindent
Fu J. et al., 2010, MNRAS, 409, 515

\noindent
Fukui Y., \& Kawamura A., 2010, ARA\&A, 48, 547

\noindent
Galliano F. et al., 2011, A\&A, 536, A88

\noindent	
Galliano F, Galametz M, Jones A. P., 2018, ARA\&A..56..673G

\noindent
Gauger A. et al., 1999, A\&A, 346, 505

\noindent
Ginolfi M. et al., 2018, MNRAS, 473,  4538

\noindent
Gjergo E. et al., 2018, MNRAS, 479, 2588

\noindent
Gratier P. et al., 2010, A\&A, 512, A68

\noindent
Hernquist L., Katz N., 1989, ApJS, 70, 419

\noindent
Higdon J. C., Lingenfelter R. E., 2005, ApJ, 628, 738

\noindent
Hirashita H., 2000, PASJ, 52, 585

\noindent
Hirashita H., Kuo T., 2011, MNRAS, 416, 1340

\noindent
Hirashita H., Aoyama S., 2019, MNRAS, 482, 2555

\noindent
Hou et al., 2017, MNRAS, 469, 870

\noindent
Hu Chia-Yu et al., 2017, MNRAS, 471, 2151

\noindent
Hu Chia-Yu et al., 2019, MNRAS, 487, 3252

\noindent
Hughes T. M. et al., 2014, A\&A, 565, A4 

\noindent
Inoue A. K., 2011, Earth, Planets and Space, 63, 1

\noindent
Israel F. P., 1997a, A\&A, 317, 65

\noindent
Israel F. P., 1997b, A\&A, 328, 471

\noindent
Jones A. P. et al., 1994, ApJ, 433, 797

\noindent
Jones A. P., 2000, JGR (Journal of Geophysical Research), 105, 10257

\noindent
Kitayama T. et al., 2004, ApJ, 613, 631

\noindent
Komugi S. et al., 2018, PASJ: Publ. Astron. Soc. Japan, 70, 48

\noindent
Kozasa et al., 2009, ASP Conference Series, Vol. 414

\noindent
Kormendy J., Kennicutt R. C., 2004, ARAA, 42, 603

\noindent
Kormendy J., 2013, In XXIII Canary Islands Winter School of Astrophysics, Secular Evolution of Galaxies eds. J. Falcon-Barroso \& J. H. Knapen (Cambridge: Cambridge University Press), p. 1

\noindent
Krumholz M. R., McKee C. F., Tumlinson J., 2009, ApJ, 693, 216

\noindent
Kuhlen M. et al., 2012, ApJ, 749, 36

\noindent
Kuo T.-M., Hirashita H., 2012, MNRAS, 424, L34

\noindent
Leitch--Devlin M. A., Williams D. A., 1985, MNRAS, 213, 295

\noindent
Leroy A. K. et al., 2007, ApJ, 658, 1027

\noindent
Leroy, A. K. et al., 2009, ApJ, 702, 352

\noindent
Leroy A. K. et al., 2011, ApJ, 737, 12

\noindent
Leroy A. K. et al., 2019, submitted

\noindent
Mattsson L., 2011, MNRAS, 414, 781

\noindent
Mckee C., 1989, Interstellar Dust: Proceedings of the 135th Symposium of the International Astronomical Union, held in Santa Clara, California, 26-30 July 1988. Edited by Louis J. Allamandola and A. G. G. M. Tielens. International Astronomical Union. Symposium no. 135, Kluwer Academic Publishers, Dordrecht, p.431

\noindent
McKinnon R., Torrey P., Vogelsberger M., 2016, MNRAS, 457, 3775

\noindent
McKinnon R., Torrey P., Vogelsberger M., 2018, MNRAS, 478, 2851

\noindent
Meixner M. et al., 2010, A\&A, 518, L71

\noindent
Micelotta E. R., Matsuura M., Sarangi A., 2018, SSRv, 214, 58
 
\noindent
Mihos J. C. et al., 2012, ApJ, 761, 186

\noindent
Mihos J. C. et al., 2013, ApJ, 762, 82

\noindent
Mihos J. C. et al., 2018, ApJ, 862, 99

\noindent
Navarro J. F., Frenk C. S.,  White S. D. M., 1996, ApJ, 462, 563

\noindent
Neto A. F. et al., 2007, MNRAS, 381, 1450

\noindent
Nozawa T., Kozasa T.,  Habe A., 2006, ApJ, 648, 435

\noindent
Oey M. S., Lamb J. B., 2012, ASP Conf. Ser. 465, Four Decades of Research on Massive Stars, ed. L. Drissen et al. (San Francisco, CA: ASP), 431

\noindent
Sandstrom K. M. et al., 2012, ApJ, 744, 20

\noindent
Sandstrom K. M. et al., 2013, ApJ, 777, 5

\noindent
Sargent B. A. et al., 2010, ApJ, 716, 878

\noindent
Savage B. D., K. R. Sembach, 1996, AR A\&A, 34, 279, 329

\noindent
Serra Diaz--Cano L., Jones A. P., 2008, A\&A, 492, 127

\noindent
Smith D. J. B. et al., 2012, MNRAS, 427, 703

\noindent
Sofia U. J., 2004,  ASP, 309, 393

\noindent
Spitzer L. J., 1978, Physical processes in the Interstellar Medium, Wiley New York, p207

\noindent
Srinivasan S. et al., 2010, A\&A, 524, A49

\noindent
Sutherland R. S., Dopita M. A., 1993, ApJS, 88, 253

\noindent
Takagi T. et al., 2010, A\&A, 514, 5

\noindent
Thornton K. et al., 1998, ApJ, 500, 95

\noindent
Tsujimoto T. et al.,  1995, MNRAS, 277, 945

\noindent
Relano M. et al., 2016, A\&A, 595, A43

\noindent
Relano M. et al., 2018, A\&A, 613, A43

\noindent
Remy--Ruyer A. et al., 2014, A\&A, 563, A31

\noindent
Roman-Duval J. et al., 2017, ApJ, 841, 72

\noindent
Rosen A.,  Bregman J. N., 1995, ApJ, 440, 634

\noindent
van den Hoek L. B., Groenewegen M. A. T., 1997, A\&AS, 123, 305

\noindent
van Dokkum P. G, Abraham R., Merritt A., 2014, ApJ, 782, L24

\noindent
Ventura P. et al., 2012a, MNRAS, 420, 1442

\noindent
Ventura P. et al., 2012b, MNRAS, 424, 2345

\noindent
Wakelam V. et al., 2017, Molecular Astrophysics, Volume 9, p. 1--36

\noindent
Walter F. et al., 2008, AJ, 136, 2563

\noindent
Yamasawa et al., 2011, ApJ, 734, 44

\noindent
Yasuda Y., Kozasa T., 2012, ApJ, 745, 159

\noindent
Yin J. et al., 2009, A\&A, 505, 497

\noindent
Yamasawa D. et al.,, 2011, ApJ, 735, 44

\noindent
Zhukovska S. et al., 2016, ApJ, 831, 147

\noindent
Zhukovska S., Henning T., 2016, POS(LCDU2013)016

\noindent
Zhukovska S., Henning T., Dobbs C., 2018, ApJ, 857, 94

\end{document}